# A Geometry-Guided Feature-Space Framework for Interpretable Unsupervised 4D-STEM Analysis of Complex Multiphase Materials

Arda Genc[1*], Daniel A. Bolden[2], Albert Nguyen[3], Daniel S. Gianola[1], Ravit Silverstein[2*]

[1] Materials Department, University of California Santa Barbara, Santa Barbara, CA, USA

[2] Department of Materials Science and Engineering, University of Florida, Gainesville, FL, USA

[3] Computer Science Department, University of California Santa Barbara, Santa Barbara, CA, USA

*Corresponding authors. ardagenc@ucsb.edu, rsilverstein@ufl.edu

## Abstract

*Rapid, high-fidelity analysis of four-dimensional scanning transmission electron microscopy (4D-STEM) datasets from heterogeneous materials remains challenging because coexisting phases, orientation variants, local strain gradients, and intrinsic TEM thin-foil effects produce complex, overlapping diffraction signatures. Here, we present a machine-learning (ML) framework that combines physics-informed geometric descriptors with weighted high-dimensional embeddings to identify latent structural patterns with minimal prior knowledge. By constructing a regularized descriptor space before embedding and density-based clustering, the method improves cluster stability, suppresses noise-driven fragmentation, and reduces sensitivity to embedding hyperparameters. Spatial weighting produced the largest stability gain, increasing median cluster persistence from 0.04 to 0.55 while reducing the noise fraction from approximately 22% to 0%. Rotation weighting increased DBCV from 0.80 to 0.92, although the noise fraction rose from 0.2% to 1.0%, highlighting a trade-off between improved density separation and noise rejection. We validate the framework on two benchmark systems: Si/* $Si_xGe_{1-x}$ *epitaxial layers with continuous lattice variation and deformation, and multiphase Ti–Nb alloys containing distinct phases and orientation variants. Comparisons with a pretrained vanilla autoencoder and model-free pattern vectorization show that the physics-informed geometric descriptor provides higher cluster fidelity, greater interpretability, and substantially lower dimensionality.*

## Introduction

4D-STEM generates vast, information-rich datasets, driven by rapid advances in high-throughput detector architectures [1]. In this modality, the electron beam records a series of diffraction patterns that encode highly correlated information about a material's phase, deformation, and orientation structure [2–6]. Data volume increases further in 5D-STEM with the addition of a temporal axis, resulting in terabytes of memory usage per user session. The key challenge is no longer data acquisition, but rapidly and reliably navigating these complex datasets and transforming them into physics-informed, readily interpretable visualizations and analyses necessary to derive deep materials inferences [1].

Current analysis pipelines remain highly parameterized, sequentially complex, computationally demanding, and domain-specific, and they often fail to resolve the latent relationships that govern material behavior. For example, the widely used template-matching approach for phase and orientation analysis requires prior crystallographic models and extensive template libraries, which become increasingly difficult to construct and search for complex multiphase systems, orientation variants, strained regions, or previously unknown phases [5,7,8]. Several 4D-STEM workflows use unsupervised ML and clustering to improve analysis while preserving the important details of large diffraction datasets [9–14]. Prior studies have combined clustering with principal component analysis (PCA) for linear dimensionality reduction, while others employ uniform manifold approximation and projection (UMAP) for nonlinear embedding and visualization [9–18].

Notable examples include latent-space mapping of atomically-resolved electric fields in Si-doped graphene from atomic-resolution STEM data [12] and a high-dimensional diffraction-vector workflow that uses guided multi-stage clustering to delineate phase, rotation, and bend contour effects in 2D materials and relatively thin TEM samples [11,17,18].

In more complex, thicker TEM samples, clustering robustness is often improved through preprocessing steps. Common strategies include masking to suppress the dominant central beam or Fourier-transform-based treatments to reduce tilt and thickness effects before flattening diffraction patterns into high-dimensional intensity descriptors [14]. More recently, energy-dispersive spectroscopy chemical maps have been incorporated to enable multimodal clustering approaches [19–22].

In 4D-STEM, physically meaningful analysis requires a structured representation that preserves both diffraction-pattern signatures and their spatial context before applying neighborhood-based learning, manifold embedding, and clustering [9,23,24]. The goal is not merely dimensionality reduction but organizing diffraction-derived descriptors into a form that makes interpretation easier and robust.

A central challenge in analyzing these large, information-rich datasets is the curse of dimensionality. For example, a single 128x128-pixel diffraction pattern can be flattened into a 16,384-dimensional feature vector, where each dimension corresponds to the intensity recorded at a single detector pixel. As the dimensionality of the feature spaces expands, as in raw diffraction pattern representations, the data become increasingly sparse, local neighborhoods lose contrast, and the distinction between near and far points weakens (Aggarwal et al., 2001; Beyer Kevin and Goldstein, 1999; Donoho, 2000). These effects reduce the effectiveness of distance-based similarity measures and nearest-neighbor searches, increase sampling requirements, and make it more difficult to construct stable and physically interpretable low-dimensional representations of the data [19,28]. At the same time, many high-dimensional datasets contain fingerprint information that can be represented in far fewer dimensions than their raw measurements suggest [29]. These limitations motivate an approach in which feature-space construction is treated as a primary part of the analysis rather than as a preprocessing detail.

In this work, we present a feature-engineering-focused machine learning framework for 4D-STEM analysis that emphasizes deliberate construction of the descriptor space before embedding and clustering. By using feature extraction to control dimensionality, the framework accelerates interpretation. It enables robust, physics-informed analysis of complex structural materials without labeled training data or prior phase assignments.

Our central premise is that the main challenge in unsupervised 4D-STEM cluster analysis is not only selecting an embedding or clustering algorithm but also designing and regularizing an interpretable feature space that preserves the underlying crystallographic relationships within the data. While existing workflows often emphasize diffraction-pattern preprocessing, black-box approaches to feature extraction, such as autoencoders, or downstream dimensionality reduction and clustering, comparatively less attention has been devoted to the physics-informed design and regularization of diffraction feature representations that preserve crystallographically meaningful information [9–16,19–21,28,30].

## Feature-space construction and embedding

The workflow begins by constructing a physically interpretable feature space, referred to here as physics-informed descriptors (PIDs), from diffraction-disc centers detected by our neural-network object-detection model (Fig. 1) [31]. Before disc detection, each diffraction pattern is log-normalized and processed using Fourier-transform-based autocorrelation. This autocorrelation step reduces the influence of thickness- and tilt-related dynamical effects, similar to the robustness sought in cepstrum-based diffraction representations and improves the Friedel symmetry of the patterns [14,31,32].

The resulting descriptor representation is then passed to unsupervised embedding and clustering. As shown in Fig. 1, each diffraction pattern is encoded using canonical triplet invariants, in-plane lattice rotation, and 2D scan coordinates. These descriptors are scaled and embedded before hierarchical density-based clustering assigns labels for 2D spatial mapping and 3D visualization. For 4D-STEM datasets, the latent space is defined by three coupled approaches: the selected diffraction features, their relative weighting, and the embedding used before density-based clustering.

Feature selection determines which physical distinctions the clustering can capture. While raw diffraction intensities preserve the full signal, they are dominated by the central beam and are sensitive to background variation and detector noise. In thicker samples, the background contribution is even larger, potentially overwhelming the clustering algorithm unless additional preprocessing is applied [11,19,24,28].

In contrast, hand-crafted descriptors such as canonical triplets of Bragg vectors — together with explicit in-plane rotation and spatial-coordinate channels — compress each pattern into a compact fingerprint, rather than entangling them as in pixel-wise intensity representations.

Spatial context is widely used in computer vision segmentation tasks to preserve image structure and geometry, including in unsupervised ML frameworks such as SLIC superpixels and autoencoders [33–35]. However, their explicit integration with phase-, strain- and rotation-sensitive descriptors in unsupervised 4D-STEM analysis remains relatively unexplored. In our framework, they provide local spatial context alongside diffraction-derived descriptors.

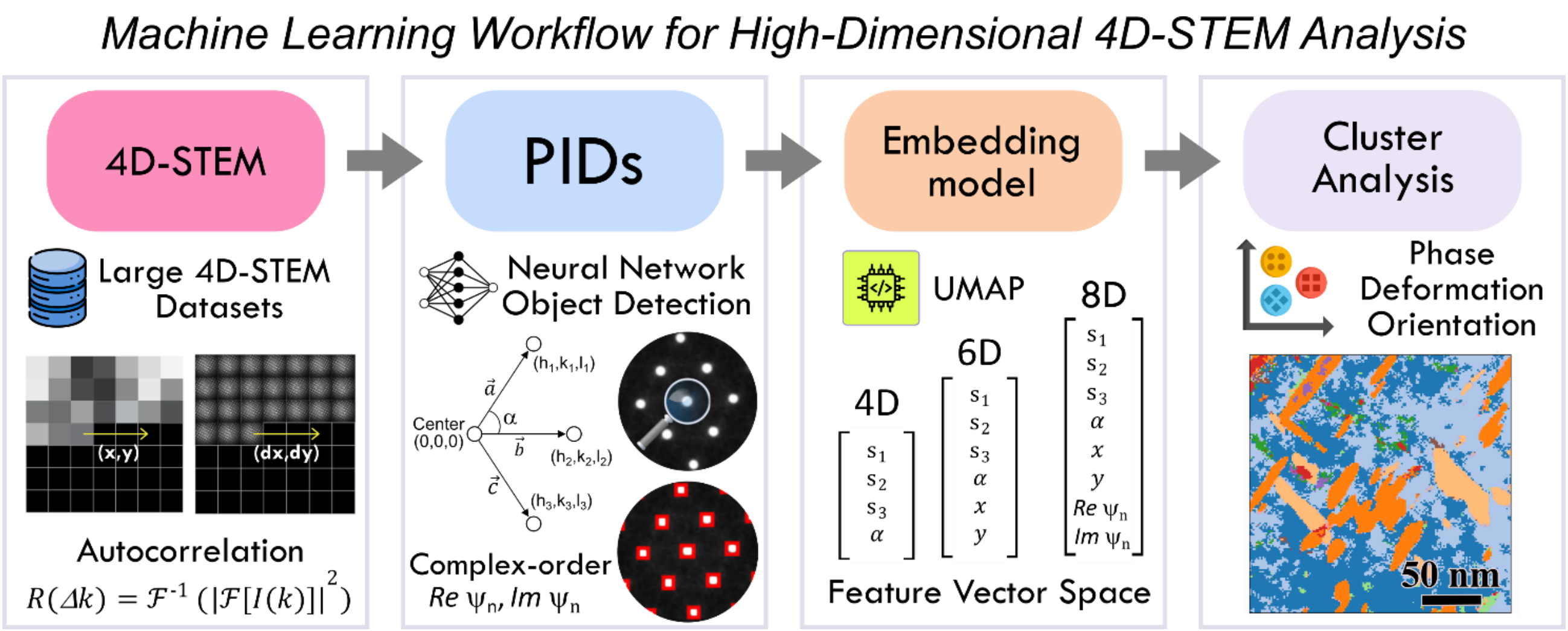

**Fig. 1.** Machine-learning workflow for high-dimensional 4D-STEM analysis. Large 4D-STEM datasets are first processed by autocorrelation and neural-network-based diffraction-feature detection. Detected disc centers are converted into physics-informed descriptors (PIDs), including canonical triplet invariants, in-plane complex orientational order parameters, and scan-position coordinates. These descriptors are scaled and embedded into a weighted feature-vector space before UMAP and density-based clustering resolve phase, deformation, and orientation structure for downstream analysis. (example: 4D-STEM dataset from a Ti-50Nb sample).

Because feature descriptors combine heterogeneous quantities, including reciprocal-space lengths, angles, and scan positions, they must be brought onto a common scale before any distance metric can be interpreted. Per-channel z-scoring removes unit imbalance, while explicit channel weights encode the relative importance of lattice geometry, rotation, and spatial coherence in the final metric. This weighted representation provides the foundation for the manifold embedding and clustering steps that follow.

To address this, the weighted feature space is embedded into a manifold that preserves local neighborhood structure while making high-density regions more accessible to the clustering algorithm. In this pipeline, feature design determines which physical attributes are represented, scaling controls the relative contribution of each channel, and embedding sets the effective resolution at which hierarchical density-based spatial clustering with noise (HDBSCAN) resolves structure [18]. Together, these steps enable pixel-resolved phase, rotation, and lattice deformation mapping from otherwise unlabeled diffraction data without requiring prior phase assignments.

## Formulation of the geometric descriptor model

For each valid pixel *i* at scan position ($x_i$, $y_i$), we construct a geometric descriptor by concatenating three feature blocks: (1) canonical triplet invariants, (2) an in-plane complex orientational order parameter, and (3) spatial scan coordinates. The corresponding notation and dimensionality for each block are summarized in Table 1.

Each feature vector is normalized, z-scored independently:

$$\tilde{x}_{\mathrm{i,c}} = \frac{(x_{\mathrm{i,c}} - \mu_c)}{\sigma_c}, \quad x \in \{g, r, p\} \tag{1}$$

Here, $i = 1, \ldots, N$ indexes the valid scan pixels, *c* the channel within each descriptor block, and $g_{\mathrm{I,c}}$, $r_{\mathrm{I,c}}$, and $p_{\mathrm{I,c}}$ the canonical triplet-geometry, in-plane rotation, and positional values at pixel *i,* as shown in Fig.1. The constants $\mu_c$ and $\sigma_c$ are the per-channel sample mean and standard deviation over the full scan, so the z-scored channels $\tilde{g}_{\mathrm{I,c}}$, $\tilde{r}_{\mathrm{I,c}}$, and $\tilde{p}_{\mathrm{i,c}}$ are dimensionless and of unit variance. The angle is the rotation-invariant interior angle $\alpha$ of the diffraction triplet, as seen in Fig. 2, calculated from $\vec{a}$, the shortest spot vector, and $\vec{b}$, the next non-collinear vector. When combined with the three magnitudes of these vectors, it yields a primitive reduced cell of the local lattice geometry in the Friedel-folded reciprocal space frame, independent of the crystal's rotation in that plane.

**Table 1.** Notation for the feature vectors, descriptors, channels, and their dimensions (D).

| Features | Descriptor | Channels | D |
|---|---|---|---|
| Geometric | $\tilde{g}_{\mathrm{i}}$ | $s_1, s_2, s_3, \alpha$ | 4 |
| Rotation | $\tilde{r}_{\mathrm{i}}$ | $Re\,\Psi_n,\ Im\,\Psi_n$ | $2\|H\|$ |
| Spatial | $\tilde{p}_{\mathrm{i}}$ | $x_i, y_i$ | 2 |

$\Psi_n$ — complex orientational order parameter

$\alpha$ — the interior angle between $s_1$ *and* $s_2$ of the primitive reduced cell

$s_1, s_2, s_3$ — the three sorted vector magnitudes (lengths of $|\vec{a}| < |\vec{b}| < |\vec{c}|$)

*H* – denotes the set of retained harmonic orders

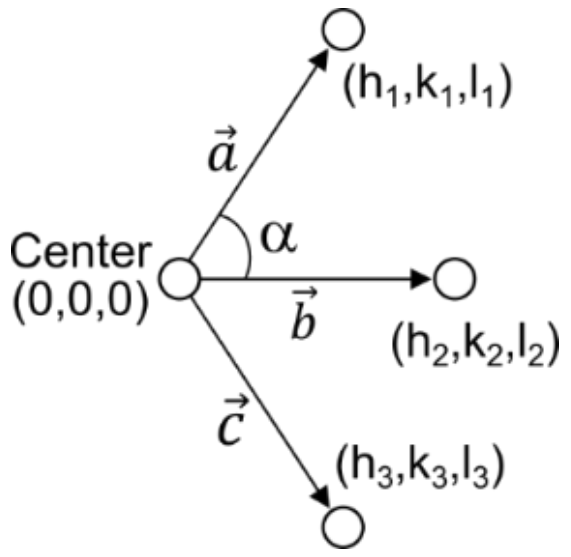

**Fig. 2.** Schematic of the canonical triplet invariant.

We define two-channel weights that encode the relative contribution of the feature axes. The rotation weight $w_r$ and spatial weight $w_p$ are simple *diagonal preconditioners* applied after z-scoring. They propagate into the Euclidean metric, analogous to inverse-covariance conditioning in Mahalanobis distance [36,37]. UMAP builds its fuzzy high-dimensional neighborhood graph from this weighted latent space.

The per-channel weights act as a diagonal preconditioner on the z-scored feature space, the variance of channel *i* scales as $w_i^2$, and the covariance between channel *i* and *j* scales as $w_i w_j$, while their Pearson correlation coefficient remains unchanged. Unlike the Mahalanobis metric, which uses inverse covariance matrix $\Sigma^{-1}$ to whiten the data and remove cross-channel correlations, this diagonal rescaling preserves the correlation structure of the descriptor and leaves the responsibility for capturing it to the subsequent UMAP embedding.

The full model vector (*Z*) that UMAP processes:

$$Z_i = \left[\, 1 \cdot \tilde{g}_{\mathrm{i,c}},\ \ w_r \cdot \tilde{r}_{\mathrm{i,c}},\ \ w_p \cdot \tilde{p}_{\mathrm{i,c}} \,\right] \qquad (2)$$

and $Z \in \mathrm{R}^{\mathrm{D}}$ where D is descriptor dimensionality. $w_r$ controls sensitivity to in-plane orientation differences, while $w_p$ controls the strength of spatial regularization by biasing the metric toward nearby scan positions.

The resulting squared weighed Euclidian distance between scan pixels *i* and *j* is defined as:

$$d^2(i,j) = d_g^2(i,j) + w_r^2\, d_r^2(i,j) + w_p^2 d_p^2(i,j) \qquad (3)$$

Here, $d_g$, $d_r$ and $d_p$ denote the Euclidean distances computed from the standardized geometric, rotational, and spatial descriptor channels, respectively, as summarized in Table 1. A linear weight in feature space becomes a *quadratic* weight in the metric.

We adapt the complex orientational order parameter $\Psi_n$ [38,39] to the azimuthal distribution of detected Bragg vectors in each diffraction pattern. Angular harmonics of scattered intensity are an established probe of local orientational symmetry in X-ray cross-correlation analysis [40] and electron diffraction of amorphous solids [41] and pattern symmetry has been used as a per-pixel contrast mechanism in STEM [42] Here, we instead evaluate $\Psi_n$ on the discrete first-order Bragg-vector ring of crystalline patterns, which yields a spot-permutation-invariant orientation descriptor.

For a pixel with N first-ring spots at azimuthal angles $\varphi_k$, the order parameter is given by:

$$\Psi_n = \frac{1}{N}\sum_{k=1}^{N} e^{(i \cdot n \cdot \varphi_k)}, \qquad n \in \{2, 4, 6\} \qquad (4)$$

Where $n$ is the rotational symmetry order, and $i$ is the imaginary unit. Fig.3 shows the geometric interpretation of the in-plane complex orientational order parameter.

We decompose the complex order parameter into real and imaginary parts, which are the two rotation channels fed to clustering after normalization. The magnitude $|\Psi_n|$ measures the symmetry strength of $n$ -fold orientational coherence, r. The full clustering descriptor combines $\mathrm{Re}\,\Psi_n$ and $\mathrm{Im}\,\Psi_n$ (Cartesian components of $\Psi_n$) with the rotation-invariant geometric triplet $[s_1, s_2, s_3, \alpha]$.

Analogous azimuthal Fourier representations of Bragg-peak positions are used in template-based automated crystal orientation mapping to solve for the in-plane rotation angle [2]; here we use the first-ring angular harmonic directly, without a reference structure, as a rotation-equivariant feature for unsupervised phase segmentation, with the dominant symmetry order selected for every diffraction pattern.

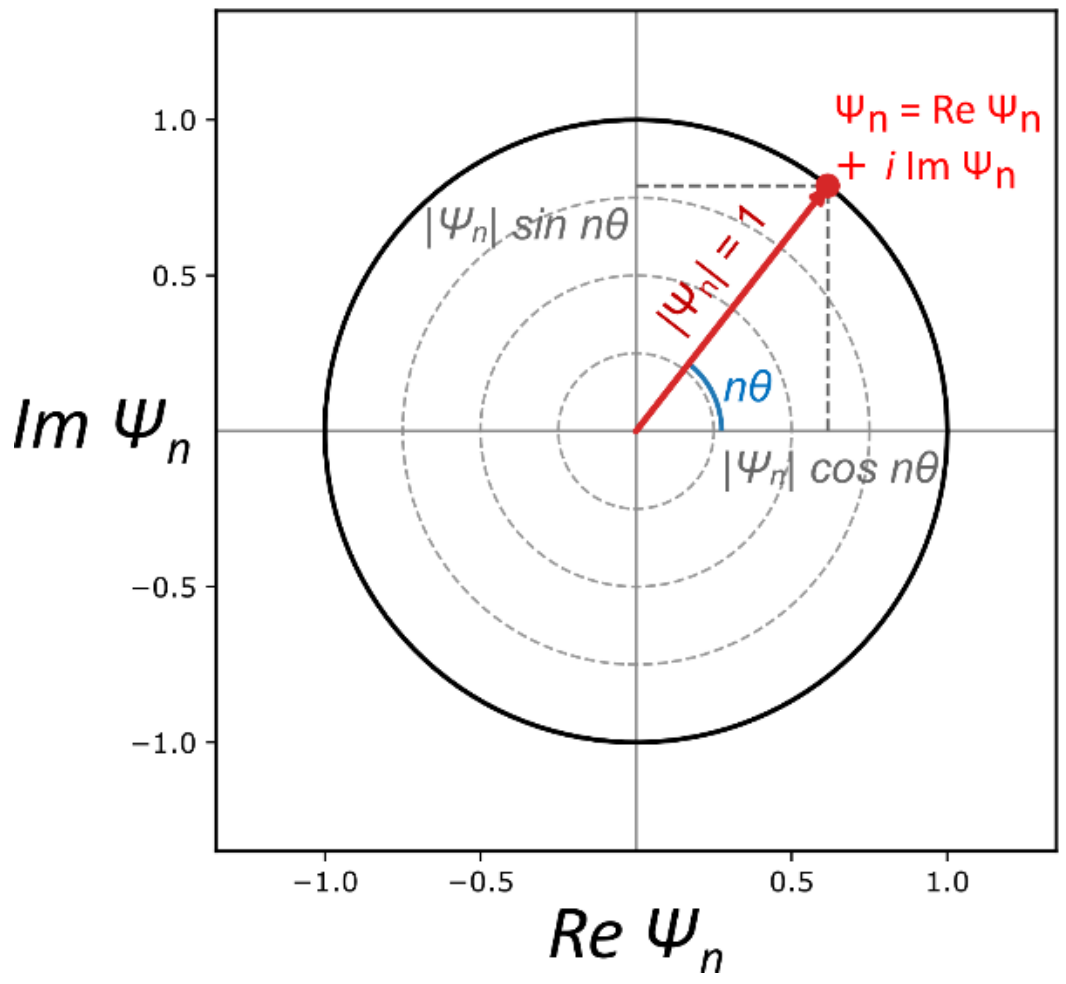

**Fig. 3.** The in-plane complex orientational order parameter $\Psi_n = |\Psi_n|e^{in\theta}$ on the complex plane- Argand diagram- (red vector). The magnitude $|\Psi_n|$ (dashed rings) is the $n$-fold coherence, r and the phase $n\theta$ (blue arc) is the lattice orientation $\theta = \arg(\Psi_n)/n$. The components $(Re\,\Psi_{n,}\ Im\,\Psi_n)$ (dashed projections) are the rotation entries of the descriptor, weighted by $w_r$.

The squared spatial contribution between two pixels separated by Δ*x*, Δ*y* scan-pixels is:

$$\mathrm{d}_\mathrm{p}^2(i,j) = \left(\frac{\Delta_x^2}{\sigma_x^2}\right) + \left(\frac{\Delta_y^2}{\sigma_y^2}\right) \approx \left(\frac{12}{H^2}\right)\left(\Delta_x^2 + \Delta_y^2\right) \quad (5)$$

For an $H \times W$ square scan, the z-scored positions have unit variance, with $\sigma_x \approx H/\sqrt{12}$ for a uniform scan. The effect of $d_p^2$ can be interpreted as a spatial coherence length. Small values impose only a weak spatial bias, with a coherence length much larger than the scan region $H \times W$, so they influence only global organization rather than local cluster identity.

As $w_p$ increases, the coherence length decreases, making spatial separation comparable to phase differences over shorter distances and causing clusters to fragment into spatially compact subregions. In this sense, the spatial weight acts as an inverse coherence length, setting how far apart two pixels can be in scan space while remaining candidate neighbors in the model metric. For each scan pixel, the descriptor channels are concatenated into a single composite feature vector in the weighted high-dimensional space, $X_{model}$. UMAP first embeds this joint representation, after which HDBSCAN computes the mutual reachability distance:

$$d_{mr}(i,j) = max\{d(i,j), d_{core}^{max}(i,j)\} \quad (6)$$

Here, $d_{core}^{max}(i,j)$ denotes the distance from $i$ to its $k$-th nearest neighbor. For 3D visualization, the high-dimensional embedding is linearly projected to 3D by PCA, retaining the three orthogonal directions of greatest variance. This projection serves only for visualization; HDBSCAN labels are assigned in the full UMAP embedding, and the 3D scatter plot renders the clustering structure in the higher-dimensional space.

## Results and Discussion

### Strain and lattice analysis of pseudomorphic Si/ $Si_xGe_{1-x}$ epitaxial layers

To evaluate the fidelity of our structural analysis workflow, we used a Si/$Si_xGe_{1-x}$ epitaxial-multilayer TEM sample (dataset I) in 4D-STEM analysis. This traceable standard consists of alternating pure Si and $Si_xGe_{1-x}$ alloy layers deposited on a single-crystal Si substrate. It contains five $Si_xGe_{1-x}$ layers, each about 10 nm thick, separated by pure Si layers about 13 nm thick, for a total superlattice thickness of approximately 100 nm. The alternating Si and $Si_xGe_{1-x}$ layers produce the light and dark contrast seen in the virtual bright-field (vBF) image in Fig. 4(a). The TEM foil thickness was approximately 70 nm, as determined by electron-energy-loss spectroscopy (EELS). We acquired XEDS-STEM profiles across the multilayer stack and measured the average Si-Ge composition using the Ge K-line at 9.874 keV and the Si K-line at 1.739 keV. The layers contained approximately 87 at. % Si (x = 87) and 13 at. % Ge. From this composition, we estimated a maximum expected out-of-plane (growth direction) strain, $\varepsilon_\perp$ , of about 0.85% using the approach of Munshi et al. [43].

The strain maps $\varepsilon_\perp$and $\varepsilon_\parallel$ in Fig. 4(b) and (c), respectively, represent the out-of-plane and in-plane (parallel to interface) strain fields, respectively, referenced to the two reciprocal-space directions used in the analysis, corresponding to the ⟨100⟩ and ⟨110⟩ reflections. The strain fields were calculated by tracking diffraction-disc centers with the center-detection model and measuring their displacements relative to a strain-free region in the Si substrate using polar decomposition (Supplementary material: polar decomposition algorithm). Oscillations in the $\varepsilon_\perp$map indicate tensile strain along the growth direction, with a mean variation of approximately 1%, in good agreement with the expected theoretical maximum for the $Si_{87}Ge_{13}$ layers. The per-pixel clusters' labels are remapped to their (x, y) probe coordinates, rendering a real-space map.

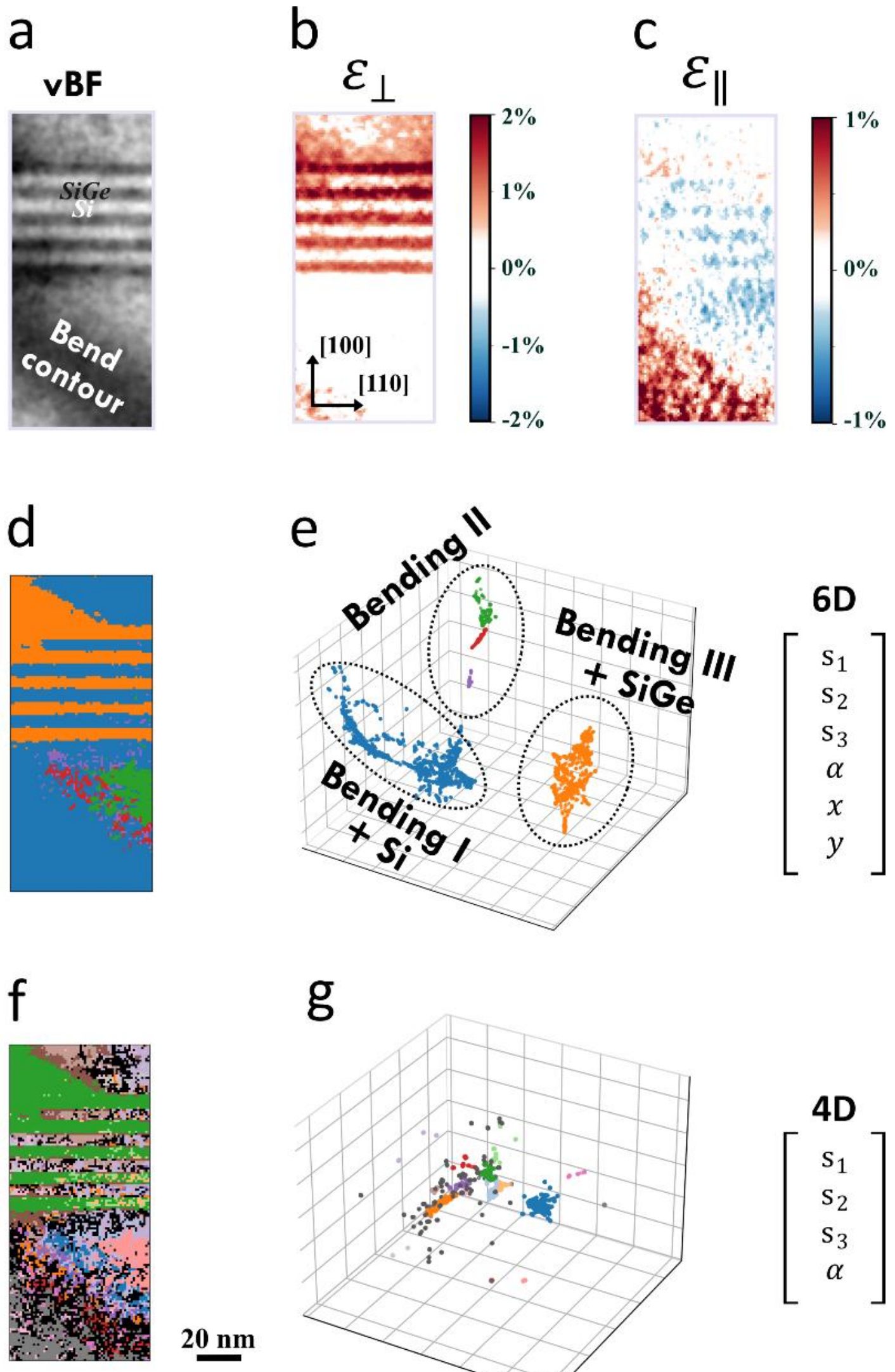

**Fig. 4.** (a) vBF image of the $Si/Si_{87}Ge_{13}$ multilayer showing layer contrast and a bend contour in the thin TEM foil. (b, c) Strain maps $\varepsilon_\perp$ and $\varepsilon_\parallel$ represent the out-of-plane and in-plane strain fields referenced to the ⟨100⟩ and ⟨110⟩ reciprocal-space directions used in the analysis, respectively. (d, e) Spatial cluster map and 3D projection highlighting foil bending and lattice variation in the Si substrate and $Si_{87}Ge_{13}$ layers ($w_p$ = 0.03) (f, g) Spatial cluster map and 3D projection with spatial weight set to zero, without any spatial regularization, showing increased noise in the spatial cluster map and fragmented clusters in the 3D UMAP embedding ($w_p$ = 0).

These contiguous regions of identical color delineate the spatial extent of each structurally distinct phase. Fig. 4(d) shows this real-space map, which we refer to as the spatial cluster map. Adjacent Fig. 4 (e) is the 3D projection of the 6D feature embedding composed of the canonical triplet invariants $s_1$, $s_2$, $s_3$, α, and the spatial coordinates $x_i$, $y_i$.

The blue and orange regions in Fig. 4(d) define a curved low-dimensional manifold in Fig. 4(e) that distinguishes Si from $Si_{87}Ge_{13}$, while also capturing a continuous ribbon-like feature associated with local lattice-spacing variations and the bend contour components labeled I, II, and III in Fig. 4(e). Collectively, these clusters reflect both Si/ $Si_{87}Ge_{13}$ contrast and intrinsic regional TEM foil bending. The manifold structure resolves the two sides of the bend contour within the Si substrate, labeled bending I and II. By contrast, it does not separate the bending III region from the $Si_{87}Ge_{13}$ layers, both of which appear as the orange cluster in the spatial cluster map and 3D projection in Fig. 4(d, e). This behavior reflects an inherent limitation of the clustering approach, which assumes that clusters occupy high-density regions separated by lower-density gaps. In this case, the continuous deformation linking the bend contour to the $Si_{87}Ge_{13}$ layers forms a smooth gradient without well-defined boundaries, thereby precluding distinct cluster partitioning.

Two principal UMAP hyperparameters are the number of neighbors and the minimum distance. The number of neighbors sets the local neighborhood size used to construct the fuzzy graph, whereas the minimum distance controls how tightly points may pack in the low-dimensional embedding. In this study, the number of neighbors was set to 100 and the minimum distance to 0.15 based on empirical optimization, as these values produced a stable manifold (Supplementary Fig 1) with well-defined boundaries. For the weighted embedding, the spatial weight was set to 0.03, imposing a modest spatial prior.

The weighted clustering resolves the lattice difference between the Si and $Si_{87}Ge_{13}$ layers and identifies the regions strained by bending. These bend contours are also visible in the virtual bright-field image as a broad alternating band of high and low intensities (Fig. 4(a)) (Supplementary Fig. 4(a-d)). Bending of the thin TEM foil is also evident in the strain maps. In the $\varepsilon_{\parallel}$ map, Fig. 4(c), it produces continuous strain modulation of about 0.1–0.2% along the nominally strain-free ⟨110⟩ direction in the Si substrate. The spatial cluster map complements the strain analysis by providing structural context, reducing the risk of misinterpreting bending-related effects.

Fig. 4(d–g) illustrates how feature-space weighting shapes the UMAP-fuzzy graph. With spatial weighting in the 6D embedding (Fig. 4d, e), the manifold remains well defined, and clustering resolves deformation and lattice variation. In the absence of this spatial prior in the 4D embedding (Fig. 4 (f, g)), the structure fragments into noise, evident as gray over-fragmentation in the 3D projection and dark noise pixels in the spatial cluster map, leaving sparse, weakly connected regions that HDBSCAN predominantly classifies as noise. These results show that weighting reshapes the UMAP topology and facilitates k-nearest neighbors (kNN) graph construction by reorganizing local neighborhoods and rescaling feature axes in high-dimensional space. As a result, HDBSCAN identifies more stable clusters.

To assess embedding behavior across the UMAP hyperparameter space, we examined 3D projections generated without any spatial regularization; these showed greater noise, weaker cluster boundaries, and reduced stability (Supplementary Fig. 2). We then evaluated the effect of spatial channel weighting by varying it over four orders of magnitude for the Si/$Si_{87}Ge_{13}$ epitaxial layers (dataset I; Table 2). Cluster stability was quantified using two complementary metrics: median per-cluster persistence ($\tilde{P}$) from the HDBSCAN condensed-tree hierarchy and the density-based cluster validity (DBCV) score [44], both defined in the methods section.

At $w_p = 0$, the geometric descriptor produced 14 candidate clusters with $\tilde{P}$ near the noise floor (0.04), indicating over-fragmentation from per-pixel descriptor noise, with approximately 22% of pixels classified as noise, consistent with Fig. 4(f, g). A modest regularization, $w_p = 0.03$, consolidated these into five density-supported clusters ($\tilde{P}$ 0.55, DBCV 0.44, noise 0%) while preserving fine continuous deformation features such as crystal bending, as shown in Fig. 4(d, e). As the spatial weight increased, these features were progressively absorbed into compact, spatially localized clusters, with a corresponding loss of manifold detail (Supplementary Fig. 3c–d). At $w_p = 0.3$, only three clusters remained, and DBCV reached its maximum value of 0.89—an apparent improvement driven by spatial-coherence smoothing rather than descriptor-based separation.

**Table 2.** Unsupervised clustering metrics for dataset I across the spatial weight $w_p$ from 0 to 1.

**Dataset I**

| $w_p$ | k | $\tilde{P}$ | DBCV |
|---|---|---|---|
| 0 | 14 | 0.04 | 0.34 |
| 0.03 | 5 | 0.55 | 0.44 |
| 0.1 | 6 | 0.44 | 0.72 |
| 0.3 | 3 | 0.54 | 0.89 |
| 1 | 3 | 0.47 | 0.78 |

At still larger weights ($w_p = 1$), the spatial channel dominated and the embedding collapsed into larger clusters (Supplementary Fig. 3(e)); both metrics declined ($\tilde{P}$ 0.47, DBCV 0.78). The two metrics peak at different operating points—$\tilde{P}$ at $w_p = 0.03$ and DBCV at $w_p = 0.3$—and their disagreement highlights the trade-off between descriptor fidelity and spatial regularization. We therefore adopt $w_p = 0.03$ as the default because it coincides with the $\tilde{P}$ peak, where descriptor information is used most effectively with the minimum regularization needed to suppress noise-driven fragmentation. The substantial gain in cluster persistence from minimal spatial regularization ($\tilde{P}$ increasing from 0.04 to 0.55 between $w_p = 0$ and $w_p = 0.03$) reflects the role of the spatial channel as a locality prior on the feature representation rather than a smoothing filter on the cluster map.

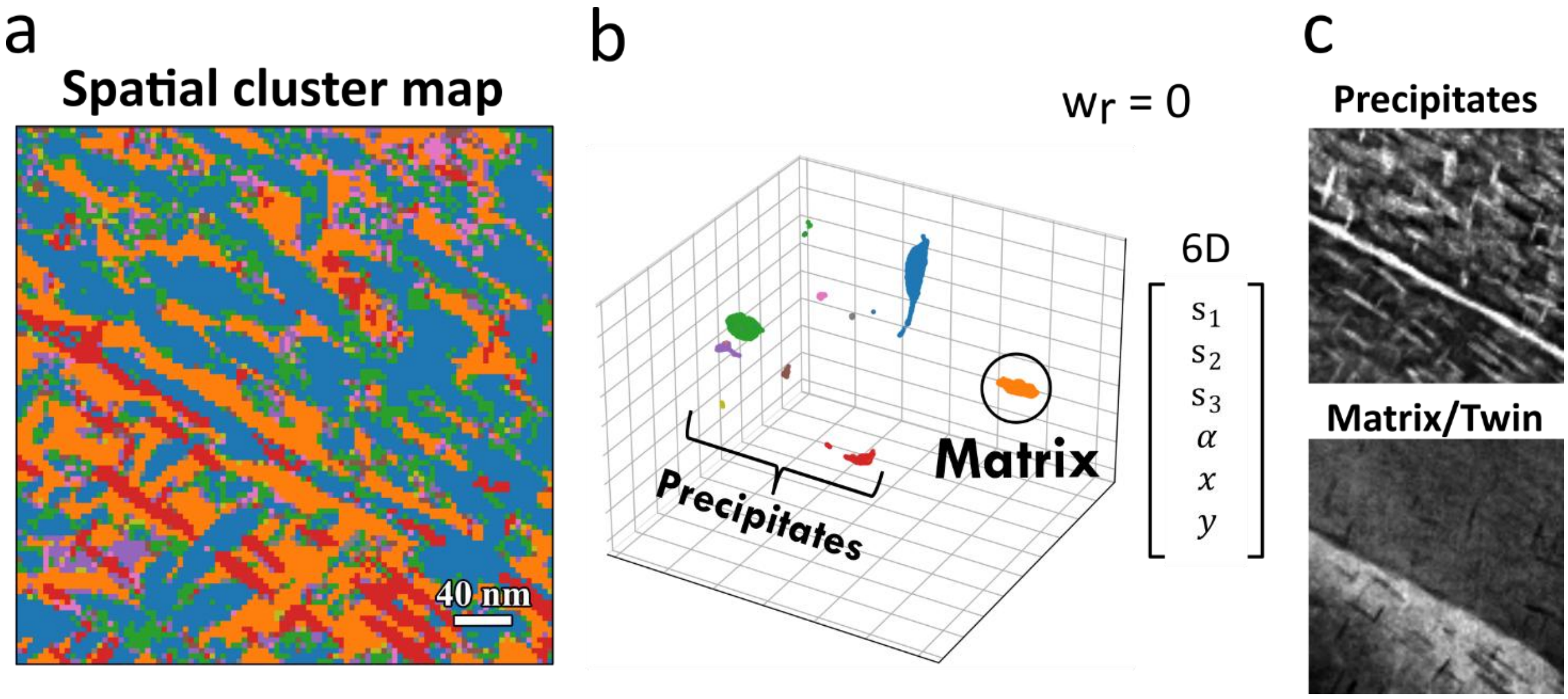

**Fig. 5.** Clustering analysis of the multiphase Ti–50Nb alloy, dataset II. (a) Spatial cluster map, showing the BCC β-Ti matrix (orange), α-HCP (red) and Ti orthorhombic α" domains (blue). (b) Corresponding 3D projection from 6D embedding (c) vDF images of the precipitates and twin matrix.

The geometric descriptor encodes lattice geometry per-pixel. Still, it discards the spatial grid structure intrinsic to 4D-STEM data, so with $w_p$ = 0, UMAP's kNN graph treats pixels at any two scan positions as interchangeable — analogous to applying a multilayer perceptron to flattened image data without the locality prior that convolutional architectures provide [44,45]. Computational performance was benchmarked on a desktop workstation equipped with an 8-core Intel i7 CPU and a 12 GB NVIDIA RTX 3060 GPU.

For the $Si/Si_{87}Ge_{13}$ dataset, the workflow processed 9,702 diffraction patterns at 512 × 512 pixel resolution, including geometric descriptor extraction and clustering, in approximately 55 s. Descriptor extraction required 45 s, corresponding to an effective throughput of about 200 frames per second, while clustering added another 10 s.

## Phase and rotation analysis in multiphase metallic alloy

We next evaluated our workflow on the more challenging task of resolving complex heterostructure in a Ti-50Nb metallic alloy (dataset II). Refractory alloys often exhibit intricate chemical variations and associated multiphase distributions together with crystallographic variants and in-plane rotations [46].

The spatial cluster map in Fig. 5(a) provides a phase-level interpretation of dataset II. Clustering is governed primarily by the rotation-invariant geometric descriptor and spatial coherence; therefore, variants with similar reduced-cell geometry remain grouped even when they differ in in-plane orientation. The orange background corresponds to the body-centered-cubic (BCC) β-Ti matrix viewed along the $[110]_\beta$ direction., while the red and blue precipitate regions correspond to hexagonal-closed-pack (α-HCP) and orthorhombic O/α" domains (distinguished by the level of tetragonal distortion of the BCC phase), respectively. In the 3D projection of the embedding in Fig. 5(b), these precipitate-related clusters are separated mainly by differences in lattice metric in Ti-based alloys [47–50]. The virtual dark-field (vDF) images in Fig. 5(c) reveal the complex phase morphology of this alloy, where needle-like and globular domains are embedded on both sides of the twinned BCC matrix, forming a highly overlapping basket-weave-like structure.

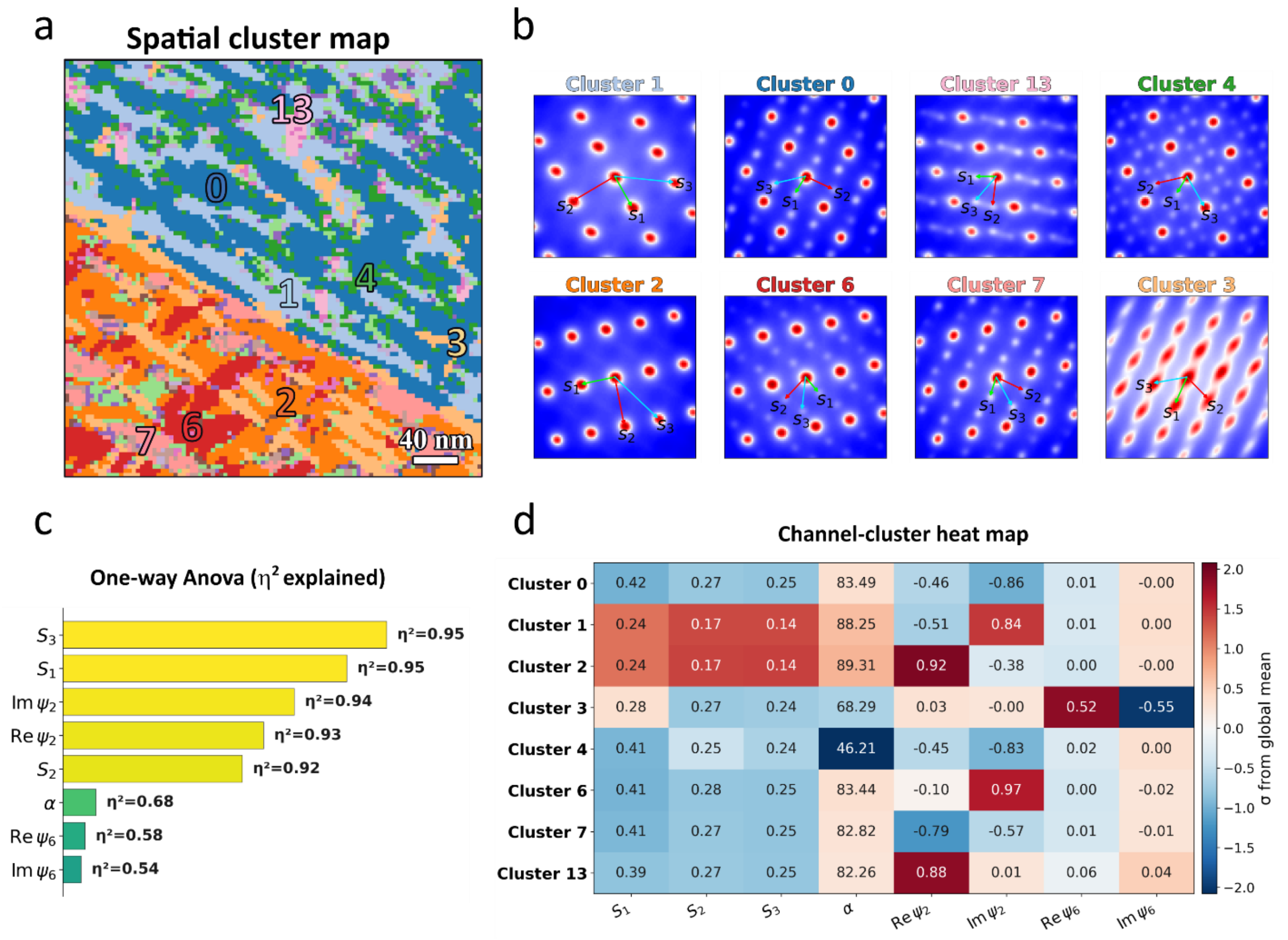

**Fig. 6.** Geometric-descriptor clustering of dataset II. (a) The spatial cluster map is shown with each cluster and (b) medoid patterns color-coded by cluster ID and annotated (green $s_1$, red $s_2$, cyan $s_3$) triplet vectors. The descriptor is robust for simple single-lattice patterns, as demonstrated by clusters corresponding to the matrix twin (Clusters 1 and 2) and precipitate variants (Clusters 0, 13, 6, 7, and 3). (c) One-way ANOVA $\eta^2$ explained (d) Channel-cluster heat map

The asynchronous object-detection workflow processed 10,000 diffraction patterns in 150 s, yielding an effective throughput of 150 fps with a batch size of 64. As in the $Si/Si_{87}Ge_{13}$ case, clustering required about 10 s, bringing the total analysis time to about 2.5 min.

The spatial cluster maps were post-processed with spatial kernel averaging (3x3) to remove noise and unassigned pixels from the UMAP embedding. This process reduces isolated wrong assignments within uniform domains while preserving grain boundaries and features, and it is applied purely as a visualization step — all reported cluster-stability and quality metrics ($\tilde{P}$, DBCV) were computed on the raw, pre-filtered HDBSCAN labels.

Incorporating the coherence-gated multi-order channels (Re $\Psi_n$, Im $\Psi_n$; n = 2, 6) into the 10-D embedding resolves three coexisting phases in a Ti-50Nb powder alloy in a single unsupervised pass (Fig. 6a,b). First, the β-Ti matrix (Clusters 1, 2): exhibits an identical BCC cell (d ≈ 0.239, 0.170, 0.139 nm — {110}, {200}, {211}; α ≈ 90°). The two clusters represent twin domains viewed along the $[110]_\beta$ zone axis, distinguished only by in-plane orientation, with a 72° rotation and $\Psi_2$ coherence r ≈ 1 (Fig. 7b). Second, an orthorhombic martensite phase, denoted α″ oriented along $[001]_{\alpha''}$ (Clusters 0, 13, 6, 7): a common large cell (d ≈ 0.41, 0.27, 0.25 nm; α ≈ 83°) appearing as four $\Psi_2$-distinct orientation variants (0/13 and 6/7, each pair ≈ 60° apart) that tile the basket-weave according to the adjoining matrix twin. The large variations in lattice spacing are likely associated with double diffraction in the TEM specimen. Third, α-HCP precipitates (Cluster 3): near-equal lattice spacings (~0.27 nm) with $\Psi_2$ incoherent (r = 0.03) but strong six-fold coherence ($|\Psi_6|$ = 0.76), the $[0001]_{\alpha\text{-HCP}}$ projection of a hexagonal grain. Note that this is a phase that a single-order $\Psi_2$ embedding cannot represent and that is recovered here only because the descriptor is influenced by $\Psi_6$. Cluster 4 traces the β/α″ phase boundaries: it inherits the $\Psi_2$ orientation of Cluster 0 but carries a distorted metric (α = 46.6°) arising from the double diffraction effect at the overlapping interfaces and is not a crystallographic distinct phase. The observed phase geometry is consistent with the orientation relationship $[110]_\beta$ // $[001]_{\alpha''}$ // $[0001]_{\alpha\text{-HCP}}$, which places the β matrix, orthorhombic α″ martensite, and α-HCP phase along aligned zone axes and provides a crystallographic basis for the variant selection observed in the cluster map. α-HCP (Cluster 3) appears as elongated laths localized almost entirely within the lower β-twin domain (Cluster 2), indicating preferential nucleation on one twin variant rather than uniform distribution — a variant-selection effect consistent with a fixed β→α orientation relationship. Here, UMAP parameters matched dataset I (100 neighbors, minimum distance 0.15); rotation and spatial weights were $w_r$ = 0.25 and $w_p$ = 0.03.

A one-way ANOVA across clusters ranks the descriptor channels by discriminability (Fig. 6 c,d). The lattice-metric edges and the $\Psi_2$ channels dominate — $s_1$ ($F$ = 24640, $\eta^2$ = 0.97), $s_3$ ($F$ = 24131, $\eta^2$ = 0.97), Im $\Psi_2$ ($F$ = 21962, $\eta^2$ = 0.97), $s_2$ ($F$ = 12914, $\eta^2$ = 0.95) and Re $\Psi_2$ ($F$ = 11964, $\eta^2$ = 0.95) — confirming that metric and dominant two-fold orientation separate the β and α″ families. The $\Psi_6$ channels rank lowest overall (Re $\Psi_6$: $F$ = 1350, $\eta^2$ = 0.67; Im $\Psi_6$: $F$ = 942, $\eta^2$ = 0.58) because only one cluster is six-fold coherent: they are the sole channels that isolate the α-HCP grain (Cluster 3), which is otherwise metric- and $\Psi_2$-ambiguous. Global channel importance and phase-resolving power are therefore distinct — a low-variance channel can still be indispensable for a minority phase, which is precisely the advantage of retaining multiple coherence-gated orders.

Complex orientational order provides a direct check that the $\Psi_2$ and $\Psi_6$ channels encode meaningful in-plane symmetries. Fig. 7 summarizes representative medoid patterns for the matrix twin, the dominant precipitate variants, and the six-fold family (Cluster 3). In the first-order ring plots (Fig. 7(a)), detected Bragg vectors are shown in reciprocal space with the direct beam at the origin; the inner solid circle marks min|g| and the outer dashed circle the $\Psi_n$ ring cutoff (30% width). Spots inside this band (filled symbols) are exactly those summed in the complex orientational order parameter — the same first-ring spots enter both the $\Psi_2$ and $\Psi_6$ sums — whereas excluded spots (hollow grey) lie on higher-order shells. The matrix twins (Clusters 1 and 2) occupy the same ring radius — confirming an identical rectangular reciprocal cell — but differ in the azimuth of their first-ring spots, i.e., in orientation only. The large-cell precipitate variants differ from the matrix by metric (a larger reciprocal cell, smaller |g|; d ≈ 0.40–0.42, 0.27, 0.25 nm, also evident in the channel-cluster heatmap of Fig. 6(d)) and from one another by orientation. Indeed, clusters 0 and 13 share this ring but differ in spot alignment, as do Clusters 6 and 7. Cluster 3, shown on its n = 6 ring, presents an approximately six-fold arrangement of first-ring spots with high coherence ($r_6 = |\Psi_6| \approx 0.76$); this is the direct-space confirmation that its separation reflects genuine six-fold orientational order rather than residual noise in the two-fold channel. The cluster-average orientation θ and coherence r (Fig. 7b) make the variant structure explicit. The β twins (Clusters 1, 2) and all four α″ variants (0, 13, 6, 7) sit at r ≈ 0.96–0.99, i.e. near-perfect $\Psi_2$ coherence, but at distinct angles: the β pair at θ = 60.5° and 168.6° (≈ 72° apart), and the α″ variants spread across θ = 47.7°, 107.9°, 121° and 180° in the ≈ 60°-spaced pairs that build the basket-weave. Cluster 3 is the diagnostic outlier — it collapses to r = 0.03 on the $\Psi_2$ dial (no two-fold orientation), and its coherence instead resides on the $\Psi_6$ order, marking it unambiguously as the hexagonal α-HCP grain rather than a poorly-oriented matrix or martensite variant.

To assess the effect of the rotation on the clustering, we varied the rotation weight $w_r$ from a rotation-descriptor free baseline ($w_r$ = 0) across $w_r$ = 0.03, 0.1, 0.25, 0.3, and 1 for dataset II, which contains both twin variants and rotation-distinct precipitates. Joint analysis of the stability metrics — cluster count k, mean cluster persistence $\tilde{P}$, DBCV, and noise fraction — together with the spatial cluster maps reveals two regimes (Table 3; Supplementary Fig. 5a–f).

**Table 3.** Unsupervised clustering metrics for dataset II as a function of the rotation weight $w_r$ (0–1): cluster count k, mean cluster persistence $\tilde{P}$, DBCV, and noise fraction.

**Dataset II**

| $w_r$ | k | $\tilde{P}$ | DBCV | Noise (%) |
|---|---|---|---|---|
| 0 | 9 | 0.63 | 0.80 | 0.2 |
| 0.03 | 10 | 0.62 | 0.92 | 1 |
| 0.1 | 12 | 0.58 | 0.88 | 0.9 |
| 0.25 | 14 | 0.56 | 0.84 | 2.3 |
| 0.3 | 14 | 0.61 | 0.87 | 4.1 |
| 1 | 17 | 0.49 | 0.84 | 3.8 |

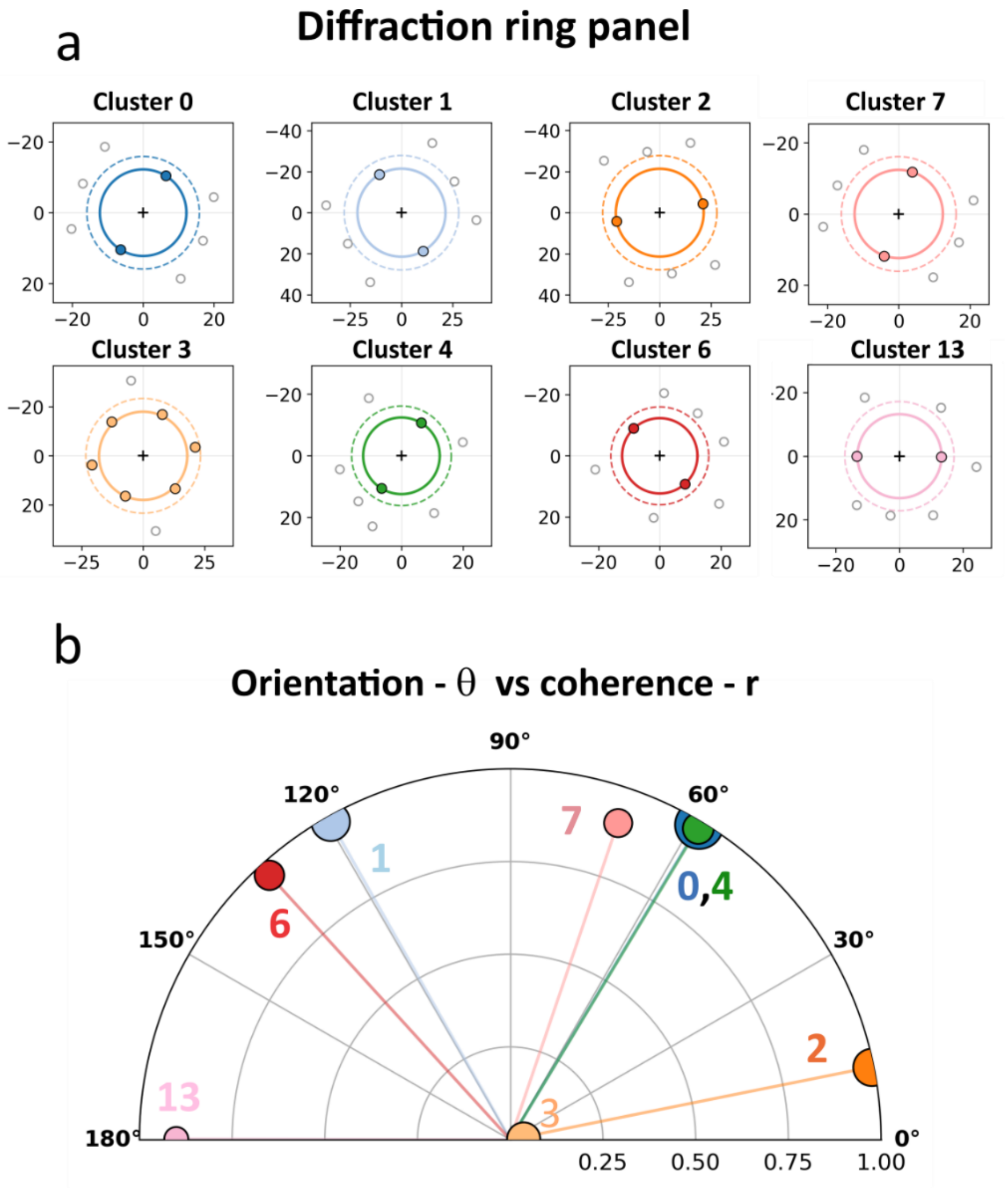

**Figure 7**. (a) Medoid diffraction rings (30% $\Psi_n$ ring width) for β matrix twins (1, 2) and orthorhombic α″domain rotational variants (0, 13, 6,7) and α-HCP (3); filled spots feed $\Psi_n$. (b) Cluster orientation θ vs coherence r polar summary (2-fold, 180° period): matrix twin ~72° separation; precipitate pairs 0/13 and 6/7 share metric but differ in orientation—colors as in Fig. 6(a).

At $w_r$ = 0, embedding yields only k = 9 clusters ($\tilde{P}$ = 0.63, DBCV = 0.80, noise = 0.2%): the metric channels alone still separate the β, α″ and α-HCP phases along the characteristic two-domain architecture (Fig. 5a), but the two β twins and the four α″ orientation variants each collapse into a single metric class, since variants that share a reduced cell are indistinguishable without orientation information. A small weight makes the embedding orientation-sensitive, and DBCV peaks at 0.92 at $w_r$ = 0.03 (k = 10), where the dominant ≈ 72° β twin is first resolved.

Increasing $w_r$ progressively resolves finer orientation variants — k = 12 (0.1) then 14 (0.25) — until the full set of β, α″ and α-HCP variants analyzed in Fig. 6(a, b) and Fig. 7 is captured at $w_r$ = 0.25 ($\tilde{P}$ = 0.56, DBCV = 0.84, noise = 2.3%). Beyond this the partition saturates: $w_r$ = 0.3 yields the same 14 clusters but nearly doubles the noise (4.1%), and only at $w_r$ = 1 does over-weighting fragment the map further (k = 17, $\tilde{P}$ = 0.49, noise = 3.8%). We therefore adopt $w_r$ = 0.25 — the smallest weight at which the complete variant set is resolved, and at the lowest noise for that partition. The marginally higher persistence and DBCV at $w_r$ = 0.3 do not indicate better structure but more aggressive noise rejection: with the same 14 clusters, the metric gain comes entirely from discarding ~1.8% more (predominantly boundary) pixels as noise. We interpret $w_r$ as an angular-resolution control, with the weight required to resolve two metric-identical variants scaling approximately as $w_r \propto 1/\Delta\theta$. The dominant β twin (Δθ ≈ 72° in-plane) is resolved at the lowest weight, where DBCV peaks ($w_r$ = 0.03). Small precipitate rotations require more weight: Clusters 0 and 7 are both α″ with an identical cell (s ≈ 0.41, 0.27, 0.25 nm; α ≈ 83°) and differ by only ≈ 13° in $\Psi_2$ orientation, so they merge into a single mode for $w_r$ < 0.25 and separate into distinct density peaks only at $w_r$ = 0.25. The rotation weight should therefore be chosen according to the angular scale of the feature of interest — coarse textures at low weight, few-degree orientation variants at higher weight — up to the point where noise degrades the partition ( $w_r \gtrsim$ 0.3).

The two datasets, dataset I (Si/ $Si_{87}Ge_{13}$) and dataset II (Ti-Nb multiphase alloy), pose fundamentally different clustering problems due to their intrinsic structures. Dataset I is dominated by continuous, non-linear lattice deformation: the per-pixel descriptors form a single, connected manifold with no intrinsic density gaps, so there is no discrete set of classes to be recovered. Clustering here is therefore better understood as a coarse-graining of a continuum than as class discovery, where the spatial weight $w_p$ acts as a granularity control, trading a finer, manifold-preserving partition for a coarser set of spatially compact regions, with the operating point chosen to match the length scale of interest. dataset II, by contrast, contains discrete phases together with twin and precipitate variants that occupy genuinely distinct, well-separated density modes. Here, clustering is well-posed and recovers physically meaningful classes: a small rotation weight ($w_r$ $= 0.03$) suffices to separate the dominant twin variant ($\sim 72°$ misorientation) while keeping the cluster density stable, whereas excessive weighting over-fragments the map. The same descriptor-and-HDBSCAN pipeline thus behaves very differently across datasets: in dataset I, it segments a continuous deformation field, while in dataset II, it resolves discrete phase and rotation variants.

**Comparison of descriptors**

Using dataset II, we benchmarked the geometric descriptor against two data-driven alternatives—a pretrained convolutional-autoencoder and a model-free flattened descriptor—using the same UMAP→HDBSCAN pipeline. The geometric descriptor produced a well-separated, spatial cluster map (6 dimensions; DBCV = 0.80, $\tilde{P}$ = 0.63), as seen in Fig. 5(a) and Table 4, whereas both data-driven descriptors failed: the 16,262-dimensional flattened descriptor was essentially unstructured (DBCV = −0.61, $\tilde{P}$=0.13), generating a continuous manifold of a single large cluster, as seen in Fig. 8(a). The 130-dimensional autoencoder resolved the dominant twin orientation by splitting the cluster into two low-density clusters, as shown in the 3D projection in Fig. 8 (c) (DBCV = -0.46, $\tilde{P} = 0.31$) and in Table 4.

The HDBSCAN condensed trees in Fig. 8 (b) and Fig. 8(d) highlight this distinction: the autoencoder and model-free descriptors produce only a few low-density modes within an otherwise continuous manifold. In contrast, the geometric descriptor yields 14 smaller clusters mostly corresponding to distinct phases, as shown in Fig. 5(a).

**Table 4.** Comparison of geometric, pretrained autoencoder, and model-free flattened descriptors for dataset II. Metrics include feature dimensionality, DBCV, median HDBSCAN cluster persistence, and noise (%). The geometric descriptor provides the highest cluster quality and stability while using the lowest-dimensional, physically interpretable representation.

| Descriptor | Dim | DBCV | $\tilde{P}$ | Noise (%) |
|---|---|---|---|---|
| Geometric | 6 | 0.80 | 0.63 | 0.2 |
| Autoencoder | 130 | -0.46 | 0.31 | 1.1 |
| Model-free | 16262 | −0.61 | 0.13 | 0.7 |

Together, these results show that the compact, physics-informed descriptor approach substantially outperforms both learned and generic high-dimensional representations. Specifically, its built-in rotation, translation, and permutation invariances isolate the discriminative lattice signal and yield genuinely discrete clusters, whereas the model-free descriptor and the autoencoder latent — optimized for reconstruction rather than class separability and pretrained on a narrow image set — encodes a relatively large continuous manifold of low cluster density. Beyond accuracy, the geometric descriptor is two-to-three orders of magnitude lower-dimensional (6 versus 130–16,262 features), making it faster to cluster and, unlike the learned and pixel-based representations, directly physically interpretable. We repeated the same descriptor comparison using dataset I, which contains simpler diffraction patterns than the multiphase Ti–50Nb alloy. At 512x512 detector resolution, the clustering results followed a similar trend: the geometric descriptor produced denser and more physically meaningful clusters with higher persistence, as shown in Supplementary Fig. 6(a– i).

Within the explored UMAP hyperparameter space, both data-driven descriptors more clearly separated deformation associated with bend contour and lattice spacing variations between the Si substrate and $Si_{87}Ge_{13}$ layers in dataset I. However, whether these latent-space separations reflect meaningful crystallographic variables rather than factors coupled with sample thickness, diffraction intensity, or beam-condition variations—remains unclear, particularly considering the strain maps in Fig.4(b, c) showing a clear strain relaxation region in Si adjacent to the top of the $Si_{87}Ge_{13}$ layer stack in the TEM sample.

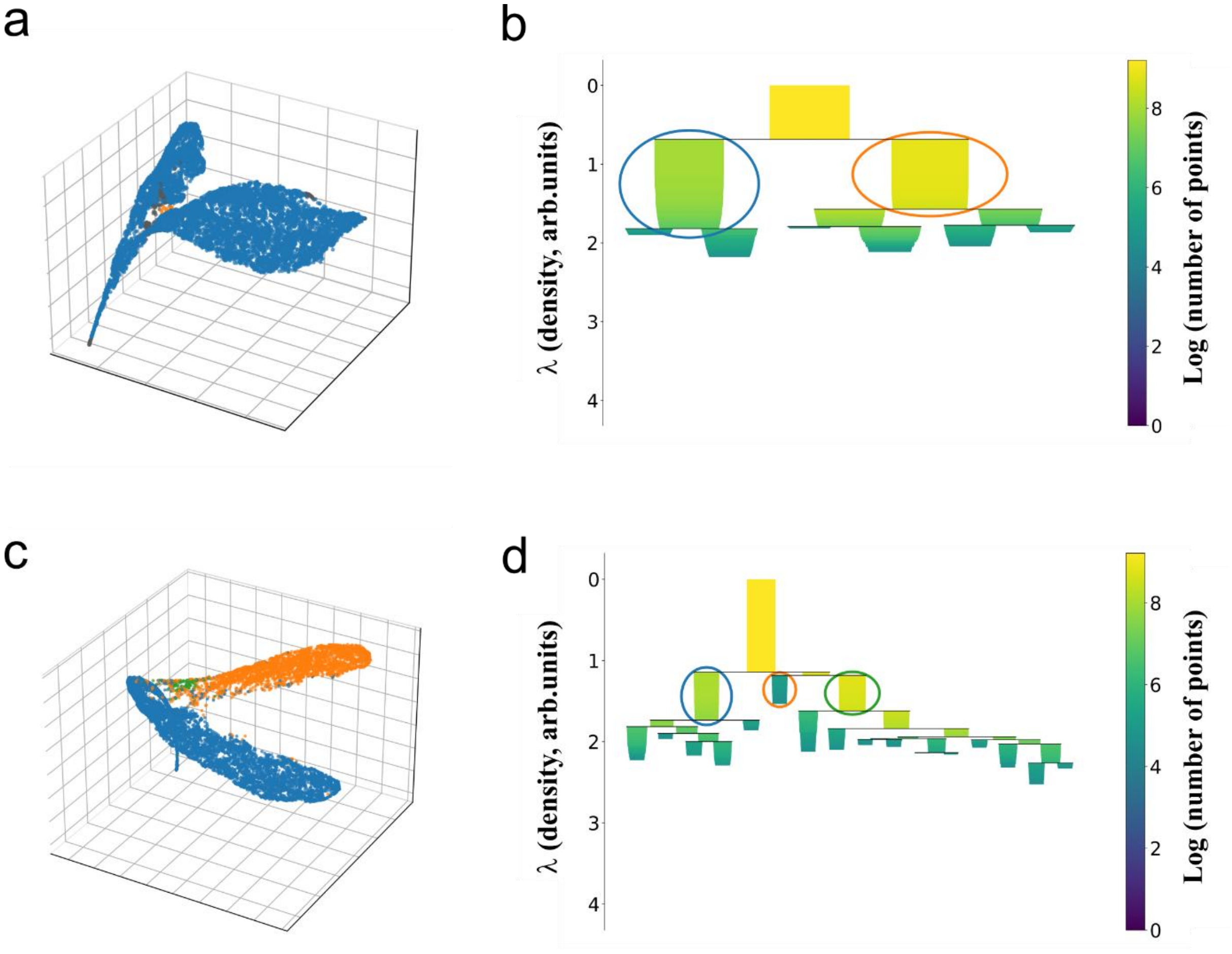

**Fig. 8.** Comparison of descriptor representations for clustering the dataset II. **(a–b)** Model-free flattened diffraction pattern descriptor results, showing primarily coarse continuous manifold clustering, like an autoencoder. **(c–d)** Pretrained autoencoder descriptor results show weaker separation of large clusters and formation of fewer stable, low-density clusters with lower persistence.

A limitation arises for high-symmetry diffraction patterns when $s_1 \approx s_2$ and $\alpha$ remains nearly constant. Over-fragmentation can occur because the clustering becomes sensitive to small numerical fluctuations rather than meaningful structural differences. Symmetry-aware canonicalization or feature weighting could reduce this instability.

In future work, we plan to expand the autoencoder architecture with residual connections to preserve spatial resolution. Additionally, we plan to curate data to expand the training datasets to include diffraction patterns acquired over a broader range of camera lengths, spot spacing, pattern geometry, and intensity distributions for both autoencoder and neural network object detection models.

## Conclusion

In this work, we present a parallelized post-processing workflow for large-scale 4D-STEM datasets that combines a neural network-based object detection model for feature extraction with feature-regularized unsupervised ML clustering. The framework enables rapid classification of diffraction features and supports accurate crystallographic analysis (assigning both phase and crystal orientation) without labeled training data or prior phase assignments. This approach proved to be successful in disentangling the various components of a complex and multiphase alloy with nanoscale features and double-diffraction effects. By accelerating interpretation across varied experimental conditions, this approach can help streamline materials discovery and understanding in systems ranging from semiconductors to metallic alloys.

## Methods

### Materials

The MAG*I*CAL® specimen was acquired from Ted Pella Inc. and is widely used to benchmark strain and lattice measurements because the lattice mismatch between epitaxial Si and $Si_xGe_{1-x}$ produces a well-defined reference system. The lattice parameters are $a_{Ge}$ = 0.5658 nm for Ge and $a_{Si}$ = 0.5431 nm for Si. A cross-section TEM sample was conventionally prepared by dimpling and Ar-ion milling.

The Ti-50Nb alloy was prepared by arc melting, followed by splat quenching to produce foils about 250 µm thick, and then homogenized to ensure a uniform bulk composition of 50 at.% Nb. Oxygen enrichment was achieved by wrapping the alloy in Nb foil, sealing it in fused silica ampoules under an $Ar/O_2$ gas mixture, and heat-treating at 800°C for 2 h to produce an oxygen concentration of about 1 at.%. TEM specimens were prepared by focused ion beam milling at 30 kV on a Thermo Fisher Scientific Helios NanoLab 600, followed by final thinning at 5 kV to minimize $Ga^+$ ion-beam damage.

### 4D-STEM data acquisition

The dataset I acquired from the Si/ $Si_xGe_{1-x}$ epitaxial-layer sample using a GATAN Stela hybrid-pixel detector operated in energy-filtered mode at 200 kV with a probe convergence angle of 0.6 mrad. The scan covered a 147 × 66 pixel region, yielding 9,702 scan positions, each recorded at 512 × 512 pixel diffraction resolution.

Dataset II was acquired from the Ti–50Nb alloy sample using a Thermo Fisher Scientific Ceta S CMOS detector and Spectra 200 STEM operated at 200 kV with a probe convergence angle of 0.6 mrad. The scan covered a 100 × 100-pixel region, with each diffraction pattern recorded at 256 × 256-pixel resolution.

### Clustering metrics

**Cluster persistence.** For cluster *k*, we define the average persistence as

$$Pers(k) = \left(\frac{1}{k}\right)\sum\nolimits_{x_i \in k}\left(\lambda_{\mathrm{death}}^{(\mathrm{k})} - \lambda_{\mathrm{birth}}^{(\mathrm{i,k})}\right) \qquad (7)$$

and report the summary statistic across clusters as

$$\tilde{P} = median\{Pers(k_i)\}_{i=1}^{K} \qquad (8)$$

- $\tilde{P}$: median persistence
- *k*: a cluster selected from the HDBSCAN condensed tree
- $|k|$: number of points assigned to cluster *k*
- $x_i$: a point belonging to cluster *k*
- $\lambda = 1 / \varepsilon$: inverse distance scale used in the HDBSCAN density hierarchy
- $\lambda_{birth}^{(i,k)}$: scale at which point $x_i$ first becomes part of cluster *k*
- $\lambda_{death}^{(k)}$: scale at which cluster *k* disappears or splits
- *K*: total number of selected clusters

**Density-based clustering validation (DBCV).** We compute the global validity score as

$$DBCV = \sum_{k=1}^{K}\left(\frac{|k|}{N}\right)V(k) \qquad (9)$$

with cluster validity term

$$V(k) = \left(\frac{DSPC(k) - DSC(k)}{\max(DSPC(k), DSC(k))}\right) \qquad (10)$$

- *N*: total number of clustered points, excluding noise points
- *V*(*k*): validity of cluster *k*
- DSC(*k*): density sparseness of cluster *k*, measuring the weakest internal density connectivity
- DSPC(*k*): density separation between cluster *k* and its nearest neighboring cluster

In this formulation, higher persistence indicates that clusters remain stable across a broader range of density thresholds. In contrast, higher DBCV indicates a cluster that is both internally dense and well separated from neighboring clusters.

**HDBSCAN condense tree visualization**

Phase clusters were identified with HDBSCAN and visualized using its condensed cluster tree. In this representation, the vertical axis is the density parameter λ = 1/d$_{mr}$, where d is the mutual-reachability distance; larger λ corresponds to higher density (smaller distance scale). Each branch ("icicle") denotes a candidate cluster, and its width represents the number of constituent points at the given λ. As λ increases, points progressively drop out of a cluster until it either splits into child clusters (when both children meet the minimum cluster size) or dissolves into noise; the persistence of a branch over a range of λ defines its stability. Branch color encodes cluster size as log (number of points), a logarithmic scale used so that clusters spanning several orders of magnitude in population remain distinguishable. The clusters retained by HDBSCAN's excess-of-mass selection are outlined and assigned the phase color palette used throughout in the spatial cluster maps.

**Channel importance (one-way ANOVA)**

To determine which descriptor channels drive clustering, we performed a per-channel one-way analysis of variance (ANOVA) across HDBSCAN cluster labels, treating each cluster as a group and excluding noise pixels (label = −1) and the scan-position channels. For channel *c* with N clustered pixels partitioned into k clusters of sizes $n_g$, we computed the between-cluster and total sums of squares:

$$SS_{between,c} = \sum_{g=1}^{k} n_g(\bar{x}_{g,c} - \bar{x}_c)^2 \tag{11}$$

$$SS_{total,c} = \sum_{i=1}^{N} (\bar{x}_{i,c} - \bar{x}_c)^2 \tag{12}$$

where $\bar{x}_{g,c}$ is the mean of channel *c* in cluster *g* and $\bar{x}_c$is the grand mean.

$$SS_{within,c} = SS_{total,c} - SS_{between,c} \tag{13}$$

From these, we report the F-ratio and effect size,

$$F_c = \frac{SS_{between,c}/(k-1)}{SS_{within,c}/(N-k)} \tag{14}$$

$$\eta_c^2 = \frac{SS_{between,c}}{SS_{total,c}} \tag{15}$$

Here $\eta_c^2$ is the fraction of each channel's variance explained by the cluster partition (0 = no separation, 1 = fully separated). Because $F_c$ and $\eta_c^2$ are invariant to constant per-channel rescaling, they are unaffected by the z-scoring and channel weighting applied before embedding; channels are ranked by $F_c$ to order their contribution to the clustering (Fig.4 (c)).

**Channel cluster heatmap**

To visualize how each cluster differs, we tabulated the mean descriptor of every cluster from the raw (unscaled) feature values. Reciprocal-vector magnitudes were converted to real-space d-spacing using the diffraction pixel size *p* (nm$^{-1}$ per pixel)

$$d = \frac{1}{s \cdot p} \tag{16}$$

With the interior angle α reported in degrees, and the complex orientational channels Re $\Psi_n$ and Im $\Psi_n$ were left dimensionless. In parallel, we computed a standardized effect for each cluster–channel pair, where $\sigma_c$ is the global standard deviation of channel *c* over all clustered pixels; this expresses each cluster mean as a deviation from the population mean in units of standard deviation, so that phase- and orientation-distinguishing channels are directly comparable across the descriptor. The resulting cluster-mean and standardized-effect tables are rendered as the channel–cluster heatmap in Fig. 4(d).

**Pretrained autoencoder descriptor.**

As a learned alternative to the hand-crafted descriptors, we used a convolutional autoencoder as a fixed feature extractor. The network is a four-stage stride-convolution encoder (32→64→128→256 channels, each Conv–BatchNorm–ReLU) feeding a 128D-channel bottleneck, mirrored by a transposed-convolution decoder; it was pretrained self-supervised (reconstruction loss) on 99 diffraction patterns — the same image domain used to train the diffraction peak center detector (neural network object detection model)—. The convolutional autoencoder was trained to minimize MSE reconstruction loss using the AdamW optimizer with an initial learning rate of 0.001, a weight decay of 0.0001, and a cosine annealing learning-rate scheduler. The model was trained for a maximum of 60 epochs with a batch size of 8. A 5% validation split and early stopping with a patience of 20 epochs were used to mitigate overfitting. The checkpoint with the lowest validation loss was retained for downstream analysis. Its weights were held fixed during clustering. At inference, each scan position's diffraction pattern was converted to its log-autocorrelation, per-image min–max normalized to [0,1], and contrast-inverted to match the polarity of the training images (bright background, dark peaks), then passed through the encoder. The bottleneck feature map was globally average-pooled and L2-normalized to yield a 128-dimensional embedding per pixel. As with the geometric descriptor, the two scan-position channels were then concatenated, yielding a 130-dimensional feature vector that was fed into the common pipeline (per-channel z-scoring, UMAP, HDBSCAN).

**Model-free descriptor**

As a model-free baseline to the geometric descriptor, each scan position was represented by its diffraction pattern autocorrelation, computed as the inverse Fourier transform of the power spectrum (Wiener–Khinchin), normalized by its central peak, and log-compressed. Each autocorrelation was area-downsampled to $N \times N$ pixels ($N = 128$), its central self-correlation peak excised with a circular mask of radius $f \cdot (N/2)$ ($f = 0.1$), and the retained pixels were max-normalized per pattern and flattened into a feature vector. These vectors were then fed to the common pipeline (per-channel z-scoring, UMAP, HDBSCAN). This descriptor is translation-invariant but rotation-variant and is therefore sensitive to in-plane lattice rotation (twin/orientation variants) as well as lattice spacing.

**DATA and CODE AVAILABILITY**

The datasets and source code generated and/or analyzed during the current study are not publicly available but are available from the corresponding author upon reasonable request, subject to any applicable institutional or collaborator agreements.

**ACKNOLEDGEMENTS**

This material is based upon work supported by the U.S. Department of Energy, Office of Science, Office of Basic Energy Sciences, as part of the Genesis Mission National Science and Technology Challenges under Award Number DE-SC0026808 AWD20795.

**AUTHOR CONTRIBUTIONS**

A.G. and R.S. acquired the 4D-STEM data. A.G. developed the machine-learning algorithm and data-analysis workflow. D.B. analyzed the Ti–50Nb alloy data, and A.N. developed and trained the autoencoder model. A.G., R.S., and D.G. edited the manuscript. D.G. supervised the experiments and analyses.

**COMPETING INTERESTS**

Authors declare no competing interests.

**ADDITIONAL INFORMATION**

**Supplementary information:** the online version contains supplemental material available at

**Supplementary information for**

# A Geometry-Guided Feature-Space Framework for Interpretable Unsupervised 4D-STEM Analysis of Complex Multiphase Materials

Arda Genc[1]*, Daniel A. Bolden[2], Albert Nguyen[3], Daniel S. Gianola[1], Ravit Silverstein[2]*

[1] Materials Department, University of California, Santa Barbara, Santa Barbara, CA, USA
[2] Department of Materials Science and Engineering, University of Florida, Gainesville, FL, USA
[3] Computer Science Department, University of California, Santa Barbara, Santa Barbara, CA, USA

*Corresponding author. ardagenc@ucsb.edu, rsilverstein@ufl.edu

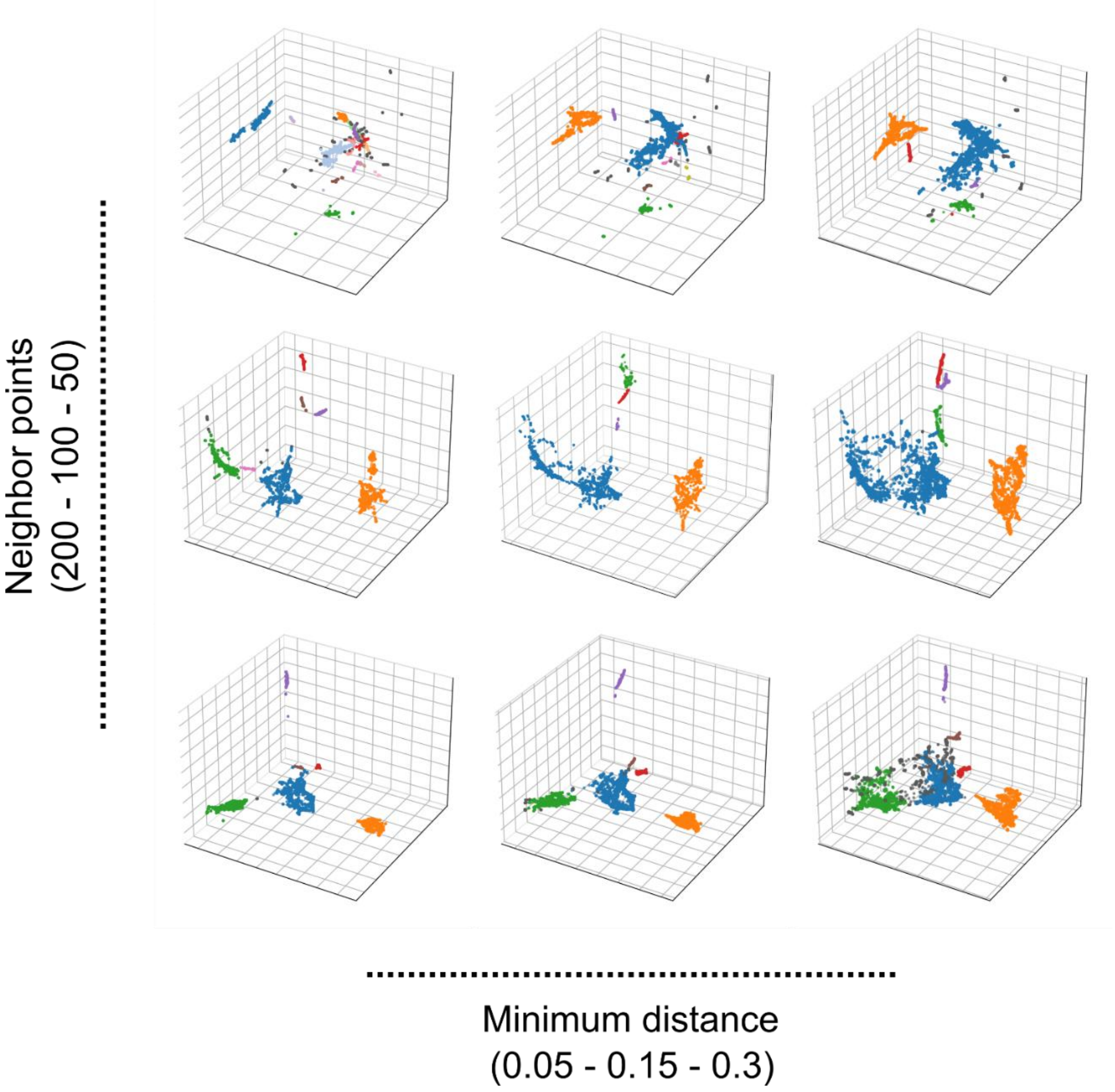

**Supplementary Fig. 1.** A series of 3D projections illustrate manifold morphology across the explored high-dimensional embedding hyperparameter space for the Si/ $Si_{87}Ge_{13}$ epitaxial-layer sample. Rows correspond to neighborhood sizes of 50, 100, and 200, and columns correspond to minimum-distance values of 0.05, 0.15, and 0.30. When spatial coordinates are included with a regularization weight ($w_p = 0.03$), the cluster structure remains largely stable across these parameter variations. Reducing the minimum distance to 0.05 at 100 neighbors decouples the bending-related contribution (green cluster) from the Si substrate (blue cluster). The most stable and well-defined cluster structure is obtained with 100 neighbors and a minimum distance of 0.15.

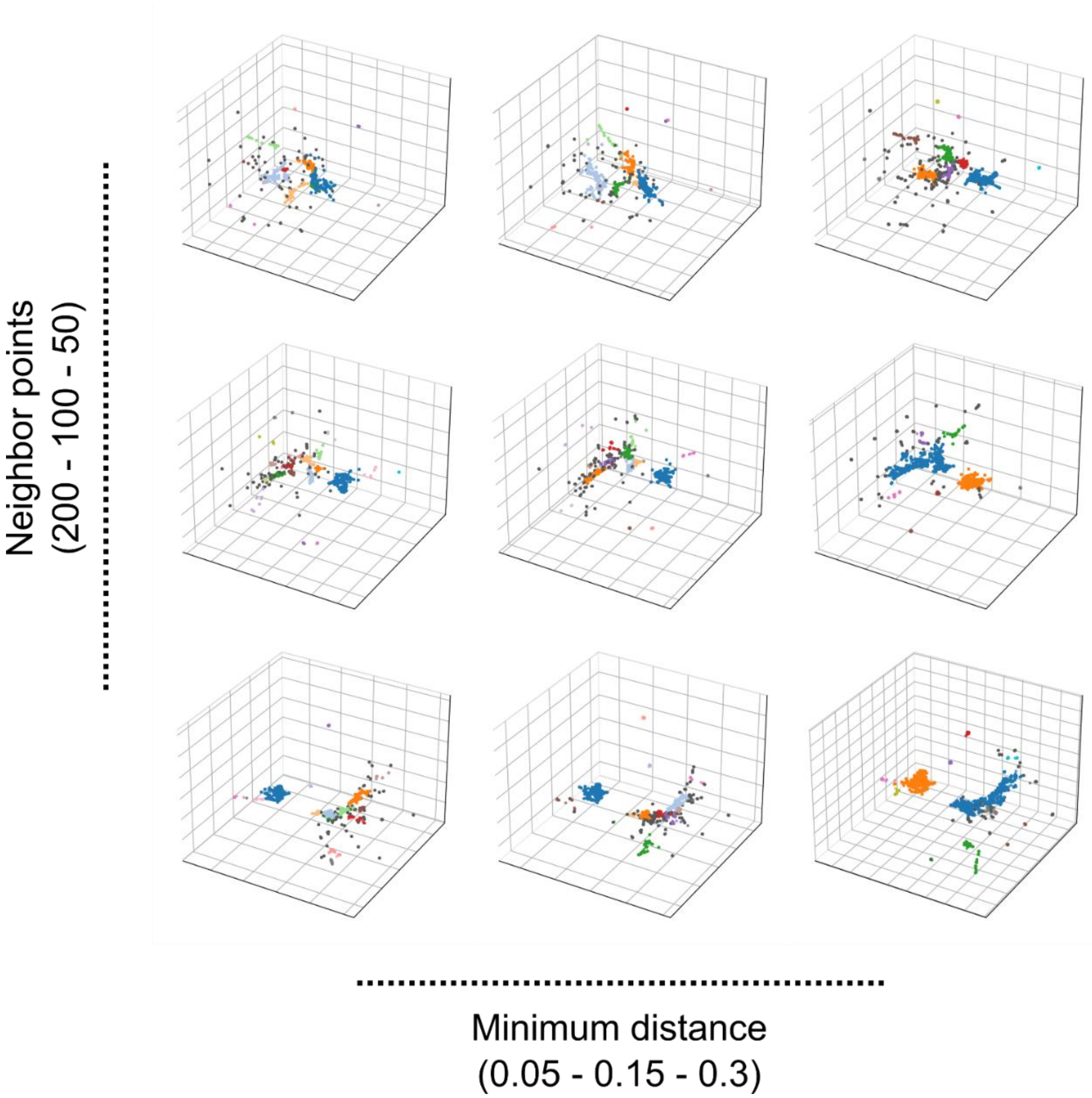

**Supplementary Fig. 2.** A series of 3D projections illustrates manifold morphology across the explored high-dimensional embedding hyperparameter space in the absence of spatial regularization ($w_p = 0$) for the Si/ $Si_{87}Ge_{13}$ epitaxial-layer sample. Rows correspond to neighborhood sizes of 50, 100, and 200, and columns correspond to minimum-distance values of 0.05, 0.15, and 0.30. Without regularization, the embedding becomes noisier and less stable across parameter variations, with weaker cluster boundaries and greater fragmentations.

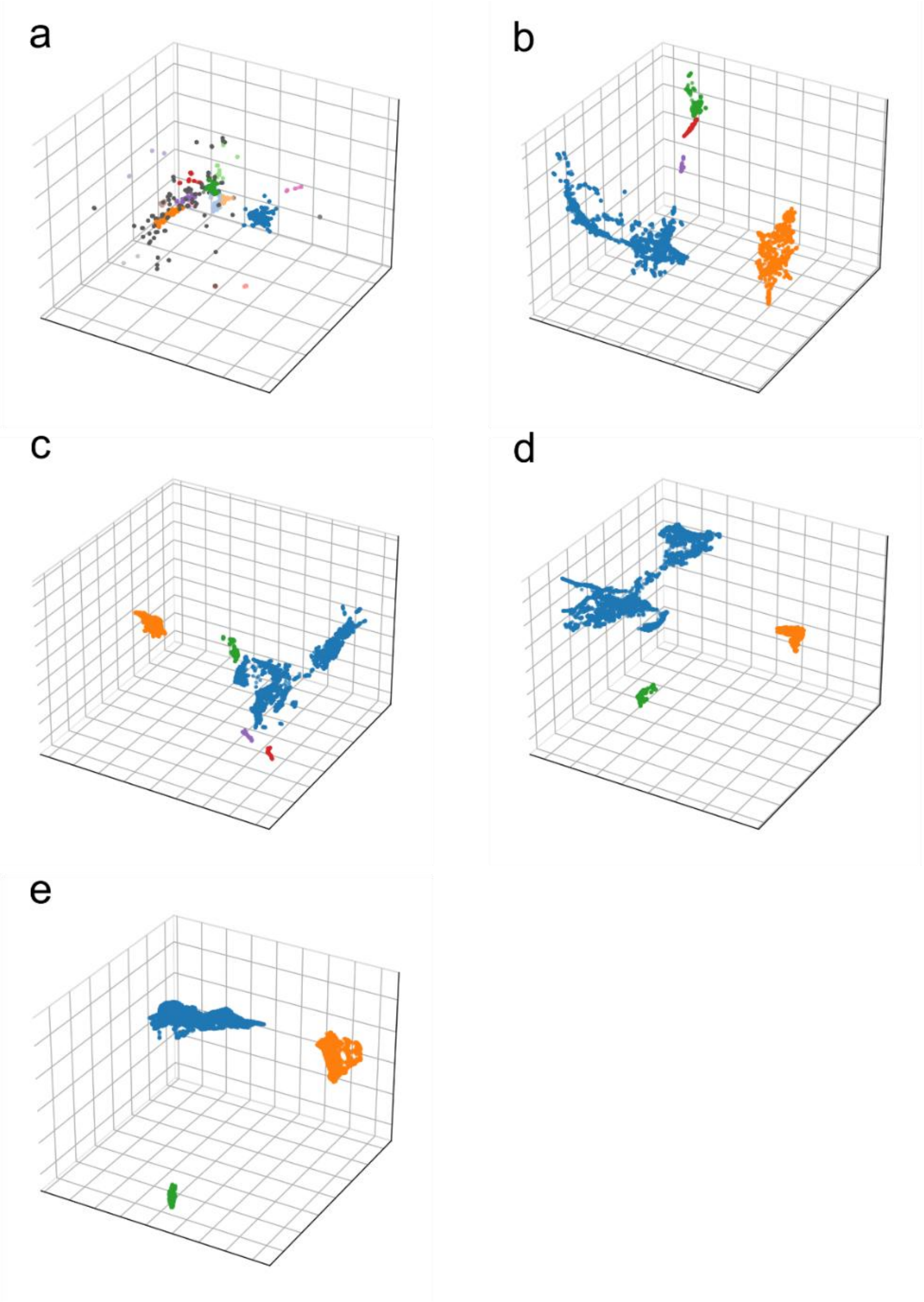

**Supplementary Fig. 3.** 3D manifold morphology across increasing spatial regularization for the Si/ $Si_{87}Ge_{13}$ epitaxial-layer sample. 3D maps **(a-e)** correspond to spatial weights ($w_p$) of 0, 0.03, 0.1, 0.3, and 1.0, respectively. As the spatial weight increases, the clusters become progressively more compact and spatially localized, with a corresponding loss of nonlinear manifold structure. This trend reflects a shift from lattice-sensitive organization toward spatially driven phase separation. The most stable and well-defined cluster structure is obtained with a spatial weight of 0.03.

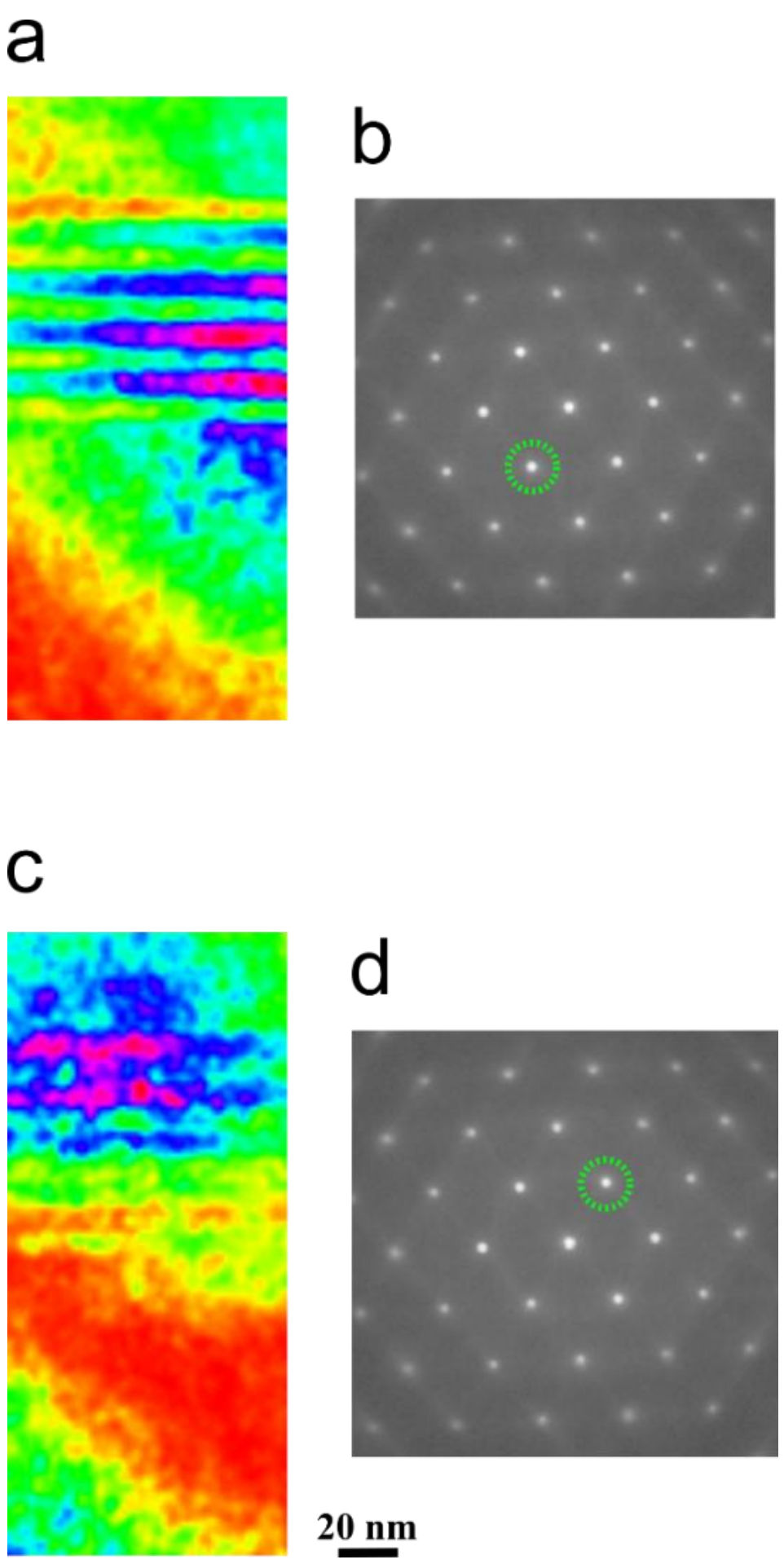

**Supplementary Fig. 4.** Schematic showing the vDF image contrast caused by the bend contour in panels **(a)** and **(b)**. Panels **(d)** and **(e)** show the virtual mask positions used to generate the corresponding vDF images in **(a)** and **(b)**.

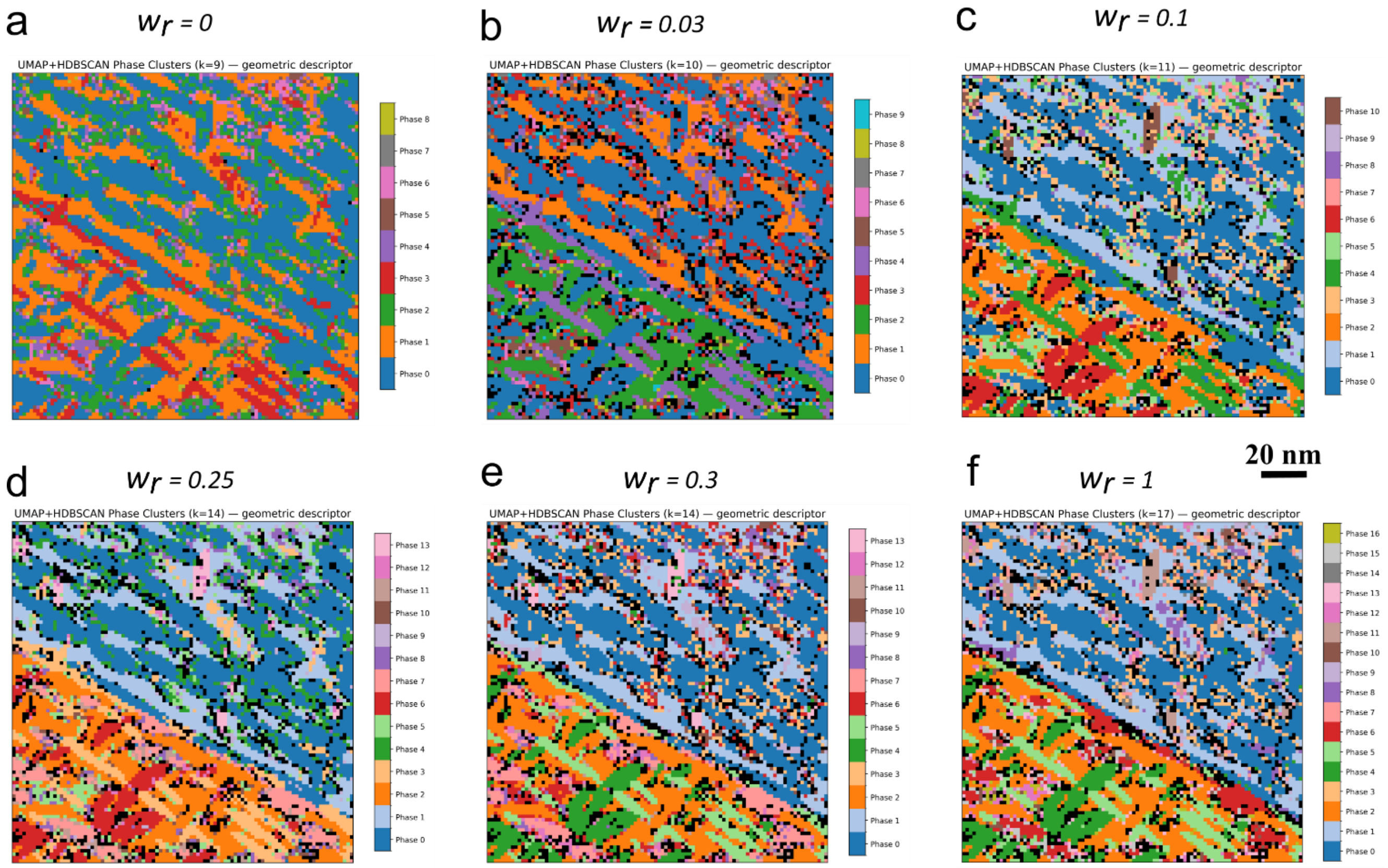

**Supplementary Fig. 5.** A series of spatial cluster maps illustrates changes in phase structure with increasing rotation regularization for Dataset II. Panels (a–f) correspond to rotation weights $w_r$ of 0, 0.03, 0.1, 0.25, 0.3, and 1, respectively. Increasing the rotation-descriptor weight progressively increases the number of clusters, cluster fragmentation, and noise, while promoting the formation of low-density clusters.

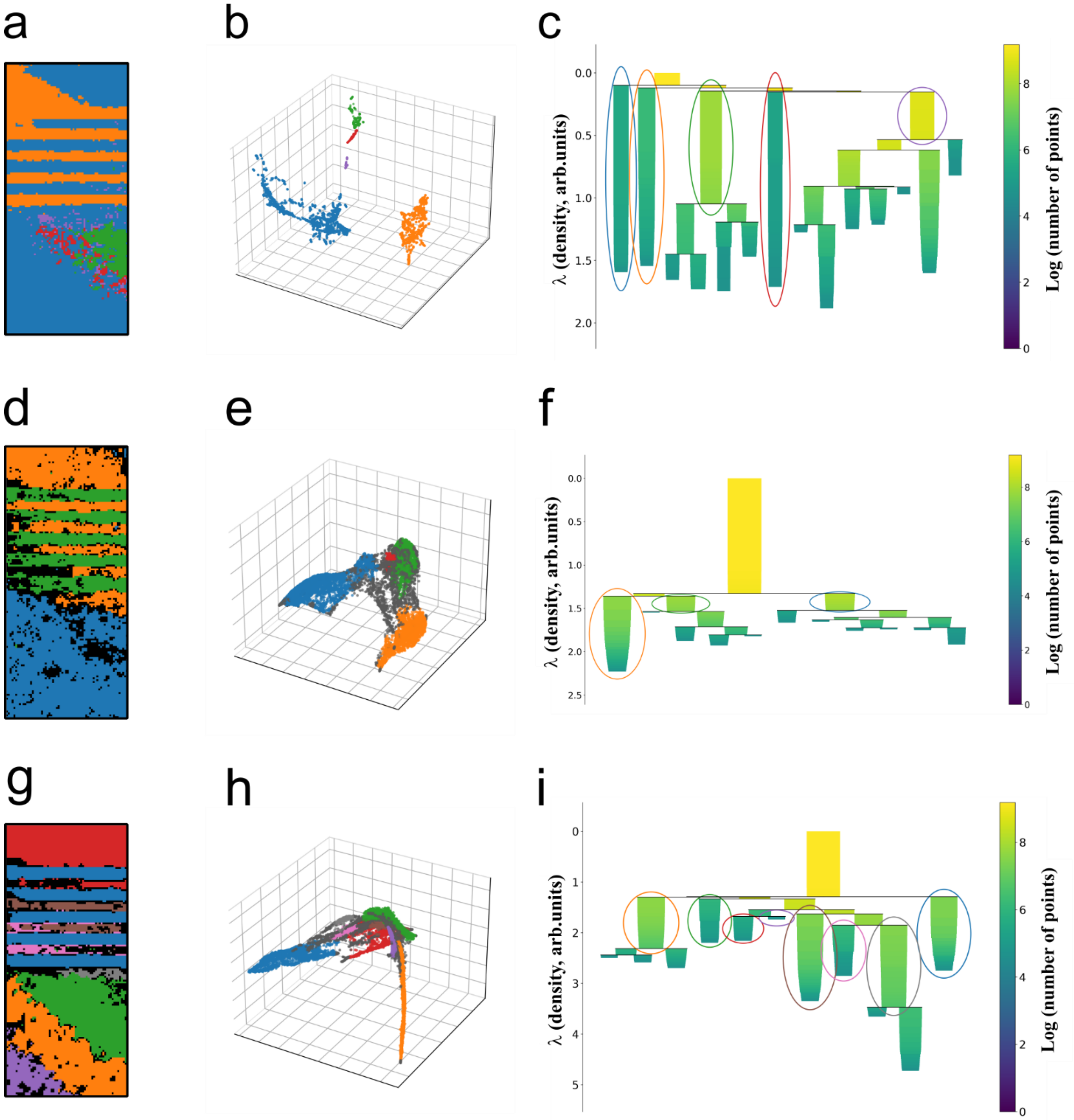

**Supplementary Fig. 6.** Comparison of descriptor representations for clustering the Si/$Si_{87}Ge_{13}$ epitaxial layer 4D-STEM dataset. **(a–c)** Geometric descriptor results, including the spatial cluster map, 3D projection, and HDBSCAN condensed tree, showing continuous manifold variation from lattice and bending deformation. **(d–f)** Pretrained autoencoder descriptor results from 128 × 128 cropped diffraction patterns, showing weaker separation and low-density large clusters which assigned as noise by HDBSCAN. **(g–i)** Model-free flattened diffraction pattern descriptor results, showing several low-density clusters and noise but identify the $Si_{87}Ge_{13}$ layers over Si background and the bend contours.

## *Polar decomposition algorithm*

For each scan position, the algorithm matches measured reciprocal-lattice vectors to a reference set, estimates the reciprocal-space transformation by least squares, converts that transformation to the real-space deformation gradient, and applies polar decomposition to separate rigid-body rotation from elastic stretch. Strain is then calculated from the symmetric stretch tensor and reported relative to the selected reference region.

**Step 1: Build the reference ($G_{ref}$) matrix**

Pick a template pixel from the ROI center. Extract its Bragg vectors — these are 2D positions of diffraction spots relative to the central beam:

$G_{ref} = \{g_1, g_2, \ldots, g_N\}$, $g_i \in$ R2

These are the "unstrained" reference vectors. Then filter them across the ROI to reduce noise and eliminate outliers.

**Step 2: Match vectors at each pixel**

At every scan pixel, the detected Bragg vectors (x, y in pixels) can be in a different order or have sign flips in Friedel pairs. At each pixel and Bragg vector, $g_j^{meas}$, compute the Euclidean distance (L2 norm) matrix against $g_i^{ref}$ considering both signs. Take the minimum and do greedy assignment (closest pair first) to establish which measured vector corresponds to which reference vector.

$d(i, j) = \| g_i^{ref} - g_j^{meas} \|_2$

*i* and *j* are the reference and measured Bragg vectors, respectively.

**Step 3: Fit the affine transformation matrix (A)**

With matched Bragg vector pairs ($g^{ref}$, $g^{meas}$), solve the least-squares problem (minimum 8 pairs from high-speed machine learning (ML) diffraction spot center detection):

$g^{meas} = A \cdot g^{ref}$

This is a 2×2 matrix A that best maps reference vectors to measured vectors in the <u>detector coordinate system.</u>

$$A = \begin{pmatrix} a_{11} & a_{12} \\ a_{21} & a_{22} \end{pmatrix}$$

$$\begin{pmatrix} x_{meas} \\ y_{meas} \end{pmatrix} = \begin{pmatrix} a_{11} & a_{12} \\ a_{21} & a_{22} \end{pmatrix} \begin{pmatrix} x_{ref} \\ y_{ref} \end{pmatrix}$$

8 pairs x 2 equations each = 16 equations for 4 unknowns (overdetermined and robust).

The fit uses at least eight matched Bragg-vector pairs, yielding at least 16 scalar equations for the four unknown entries of A. This redundancy improves numerical stability, although robustness still depends on vector geometry, localization uncertainty, correspondence quality, and outlier rejection.

**Step 4: Deformation gradient (F)**

The affine map A operates in reciprocal space. The real-space deformation gradient, *F,* is:

$F = (A^{-1})^T$

Inversion converts 'reciprocal space mapping" to "real space mapping". Transpose accounts for the difference between how row vectors and column vectors transform.

**Step 5: Polar decomposition**

Decompose F into rotation, R and stretch, U:

$F = R \cdot U$

where:

- **R** is rotation matrix (det = 1, orthogonal)
- **U** is a stretch tensor

R and U are calculated through the Numpy Singular Value Decomposition (SVD) function. The output is rotation1, stretch, and rotation2.

(1) $U_{svd}$, S, $V_t$ = np.linalg.svd(F)

(2) R = $U_{svd}$ @ $V_t$

(3) U = $V_t^T$ @ np.diag(S) @ $V_t$

Linear strain tensor (Biot strain tensor) *E*,

(4) $E = U - I = \begin{pmatrix} \varepsilon_{xx} & \varepsilon_{xy} \\ \varepsilon_{yx} & \varepsilon_{yy} \end{pmatrix}$

where :

$$I = \begin{pmatrix} 1 & 0 \\ 0 & 1 \end{pmatrix}$$

- $\varepsilon_{xx}$: normal strain in detector x-direction

- $\varepsilon_{yy}$: normal strain in detector y-direction
- $\varepsilon_{xy}$, $\varepsilon_{yx}$ : shear strain (symmetric, off-diagonal)

**Step 7: Crystal frame projection of strain tensor *E*,**

Detector frame:

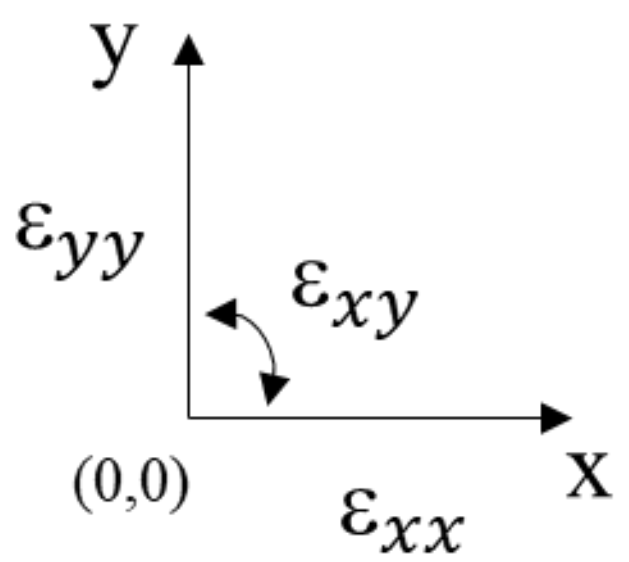

$$E = \begin{pmatrix} \varepsilon_{xx} & \varepsilon_{xy} \\ \varepsilon_{yx} & \varepsilon_{yy} \end{pmatrix}$$

$\varepsilon_{xy}$, shear strain representing the change in angle between x and y detector directions.

Crystal frame:

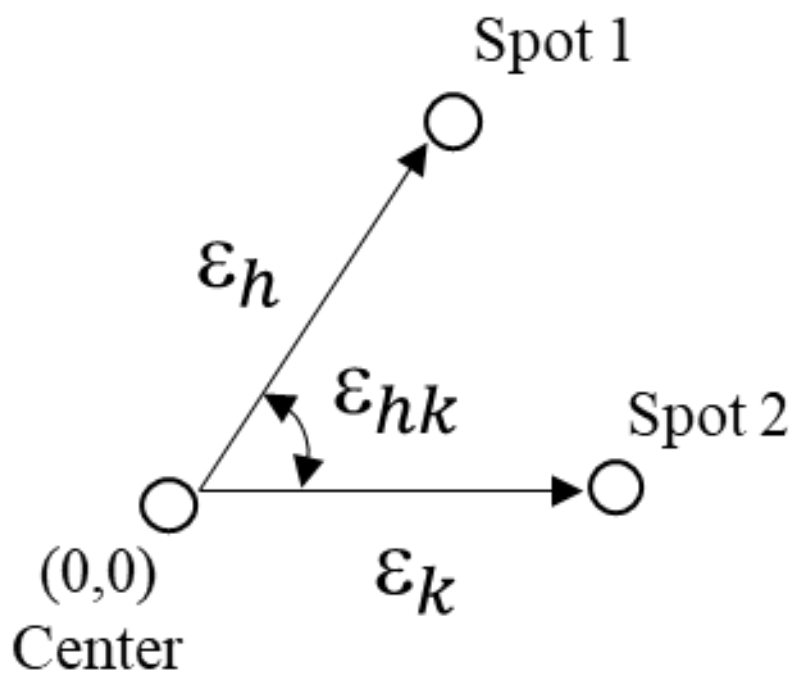

$$E = \begin{pmatrix} \varepsilon_{h} & \varepsilon_{hk} \\ \varepsilon_{kh} & \varepsilon_{k} \end{pmatrix}$$

If h and k directions are defined by the lattice directions in the diffraction pattern, to rotate the strain tensor in the detector frame to the crystal frame, we do the following:

$$\varepsilon_n = \hat{n}^T E \hat{n}$$

$$\varepsilon_h = n_x^2 \varepsilon_{xx} + 2 n_x n_y \varepsilon_{xy} + n_y^2 \varepsilon_{yy}$$

$n_i$, unit vector along the lattice direction.

$\varepsilon_h$, $\varepsilon_k$ normal strain along the lattice direction.$\alpha$

$\varepsilon_{hk}$, shear strain representing the change in angle between basis vectors.

$$\varepsilon_{xy}^{pred} = \left|\frac{(\varepsilon_1 - \varepsilon_2)}{2\ x\ sin2\alpha}\right|$$

$\alpha$ is the angle in degrees between the principal strain (no shear) and the crystal frame coordinates. $\varepsilon_1$ and $\varepsilon_2$ are rotation-free principal strain values, defining maximum and minimum strains in orthogonal directions.

**Step 8: Rotation**

$$R = F \cdot U^{-1}$$

$$R = \begin{pmatrix} R_{00} & R_{01} \\ R_{10} & R_{11} \end{pmatrix} = \begin{pmatrix} cos\theta & -sin\theta \\ sin\theta & cos\theta \end{pmatrix}$$

$$\Theta = \text{arctan2}(sin\theta, cos\theta)$$

This gives the in-plane rigid-body rotation at each pixel, relative to the ROI average orientation. Reported in degrees.

**Summary of outputs:**

| **Map** | **Meaning** |
|---|---|
| Strain A ($\varepsilon_{xx}$, $\varepsilon_h$) | Normal strain along the x detector or h crystal directions |
| Strain B ($\varepsilon_{yy}$, $\varepsilon_k$) | Normal strain along the y detector or k crystal directions |
| Strain C ($\varepsilon_{xy}$, $\varepsilon_{hk}$) | Shear strain along detector and crystal directions |
| Rotation | Rigid-body in-plane rotation (degrees) in polar coordinates |